\documentstyle[]{mn}
\newif\ifAMStwofonts

\def\simgt{\mathrel{\spose{\lower 3pt\hbox{$\sim$}}
        \raise 2.0pt\hbox{$>$}}}
\def\simlt{\mathrel{\spose{\lower 3pt\hbox{$\sim$}}\raise 2.0pt\hbox{$<$}}}

\ifoldfss
  \ifCUPmtlplainloaded \else
    \NewTextAlphabet{textbfit} {cmbxti10} {}
    \NewTextAlphabet{textbfss} {cmssbx10} {}
    \NewMathAlphabet{mathbfit} {cmbxti10} {} 
    \NewMathAlphabet{mathbfss} {cmssbx10} {} 
  \fi
  \ifAMStwofonts
    \ifCUPmtlplainloaded \else
      \NewSymbolFont{upmath} {eurm10}
      \NewSymbolFont{AMSa} {msam10}
      \NewMathSymbol{\upi}     {0}{upmath}{19}
      \NewMathSymbol{\umu}     {0}{upmath}{16}
      \NewMathSymbol{\upartial}{0}{upmath}{40}
      \NewMathSymbol{\leqslant}{3}{AMSa}{36}
      \NewMathSymbol{\geqslant}{3}{AMSa}{3E}

      \let\geq=\geqslant 
    \fi
  \fi
\fi 

\ifnfssone
  \newmathalphabet{\mathit}
  \addtoversion{normal}{\mathit}{cmr}{m}{it}
  \addtoversion{bold}{\mathit}{cmr}{bx}{it}
  \newmathalphabet{\mathbfit} 
  \addtoversion{normal}{\mathbfit}{cmr}{bx}{it}
  \addtoversion{bold}{\mathbfit}{cmr}{bx}{it}
  \newmathalphabet{\mathbfss} 
  \addtoversion{normal}{\mathbfss}{cmss}{bx}{n}
  \addtoversion{bold}{\mathbfss}{cmss}{bx}{n}
  \ifAMStwofonts
    \ifCUPmtlplainloaded \else
      \UseAMStwoboldmath
      \makeatletter
      \new@mathgroup\upmath@group
      \define@mathgroup\mv@normal\upmath@group{eur}{m}{n}
      \define@mathgroup\mv@bold\upmath@group{eur}{b}{n}
      \edef\UPM{\hexnumber\upmath@group}
      \new@mathgroup\amsa@group
      \define@mathgroup\mv@normal\amsa@group{msa}{m}{n}
      \define@mathgroup\mv@bold\amsa@group{msa}{m}{n}
      \edef\AMSa{\hexnumber\amsa@group}
      \makeatother
      \mathchardef\upi="0\UPM19
      \mathchardef\umu="0\UPM16
      \mathchardef\upartial="0\UPM40
      \mathchardef\leqslant="3\AMSa36
      \mathchardef\geqslant="3\AMSa3E

      \let\geq=\geqslant 
    \fi
  \fi
\fi 

\ifnfsstwo
  \DeclareMathAlphabet{\mathbfit}{OT1}{cmr}{bx}{it}
  \SetMathAlphabet\mathbfit{bold}{OT1}{cmr}{bx}{it}
  \DeclareMathAlphabet{\mathbfss}{OT1}{cmss}{bx}{n}
  \SetMathAlphabet\mathbfss{bold}{OT1}{cmss}{bx}{n}
  \ifAMStwofonts
    \ifCUPmtlplainloaded \else
      \DeclareSymbolFont{UPM}{U}{eur}{m}{n}
      \SetSymbolFont{UPM}{bold}{U}{eur}{b}{n}
      \DeclareSymbolFont{AMSa}{U}{msa}{m}{n}
      \DeclareMathSymbol{\upi}{0}{UPM}{"19}
      \DeclareMathSymbol{\umu}{0}{UPM}{"16}
      \DeclareMathSymbol{\upartial}{0}{UPM}{"40}
      \DeclareMathSymbol{\leqslant}{3}{AMSa}{"36}
      \DeclareMathSymbol{\geqslant}{3}{AMSa}{"3E}

      \let\geq=\geqslant 
    \fi
  \fi
\fi 

\ifCUPmtlplainloaded \else
  \ifAMStwofonts \else 
    \def\upi{\pi}
    \def\umu{\mu}
    \def\upartial{\partial}
  \fi
\fi

\title[The Log-Quadratic $M_{\rm bh}-\sigma$ Relation]{A Log-Quadratic Relation Between the Nuclear Black-Hole Masses and Velocity Dispersions of Galaxies}

\author[Wyithe]{J. Stuart B. Wyithe\\
School of Physics, University of Melbourne, Parkville, Victoria, Australia\\
 Email: swyithe@physics.ph.unimelb.edu.au}

\date{Accepted Received}
\pagerange{\pageref{firstpage}--\pageref{lastpage}}
\pubyear{2005}

\def\LaTeX{L\kern-.36em\raise.3ex\hbox{a}\kern-.15em
    T\kern-.1667em\lower.7ex\hbox{E}\kern-.125emX}

\begin{document}

\label{firstpage}

\maketitle

\begin{abstract}
\noindent 
We demonstrate that a log-linear relation does not provide an adequate
description of the correlation between the masses of Super-Massive
Black-Holes (SMBH, $M_{\rm bh}$) and the velocity dispersions of their
host spheroid ($\sigma$). An unknown relation between $\log{M_{\rm
bh}}$ and $\log{\sigma}$ may be expanded to second order to obtain a
log-quadratic relation of the form $\log{(M_{\rm
bh})}=\alpha+\beta\log{(\sigma/200\mbox{km\,s}^{-1})}+\beta_2[\log{(\sigma/200\mbox{km\,s}^{-1})}]^2$.
We perform a Bayesian analysis using the Local sample described in
Tremaine et al.~(2002), and solve for $\beta$, $\beta_2$ and $\alpha$,
in addition to the intrinsic scatter ($\delta$). We find unbiased
parameter estimates of $\beta=4.2\pm0.37$, $\beta_2=1.6\pm1.3$ and
$\delta= 0.275\pm0.05$.  At the 80\% level the $M_{\rm bh}-\sigma$
relation does not follow a uniform power-law. Indeed, over the
velocity range 70km/s$\la\sigma\la$380km/s the logarithmic slope
$d\log{M_{\rm bh}}/d\log{\sigma}$ of the best fit relation varies
between 2.7 and 5.1, which should be compared with a power-law
estimate of $4.02\pm0.33$. The addition of the 14 galaxies with
reverberation SMBH masses and measured velocity dispersions (Onken et
al.~2004) to the Local SMBH sample leads to a log-quadratic relation
with the same best fit as the Local sample. Furthermore, assuming no systematic offset, single
epoch virial SMBH masses estimated for AGN (Barth et al.~2005) follow
the same log-quadratic $M_{\rm bh}-\sigma$ relation as the Local
sample, but extend it downward in mass by an order of magnitude. The
log-quadratic term in the $M_{\rm bh}-\sigma$ relation has a
significant effect on estimates of the local SMBH mass function at
$M_{\rm bh}\ga10^9M_\odot$, leading to densities of SMBHs with $M_{\rm
bh}\ga10^{10}M_\odot$ that are several orders of magnitude larger than
inferred from a log-linear $M_{\rm bh}-\sigma$ relation.  We also
estimate unbiased parameters for the SMBH-bulge mass relation using
the sample assembled by Haering
\& Rix~(2004). With a parameterization $\log{(M_{\rm bh})}=\alpha_{\rm
bulge}+\beta_{\rm bulge}\log{(M_{\rm
bulge}/10^{11}M_\odot)}+\beta_{\rm 2,bulge}[\log{(M_{\rm
bulge}/10^{11}M_\odot)}]^2$, we find $\beta_{\rm
bulge}=1.15\pm0.18$ and $\beta_{\rm 2,bulge}=0.12\pm0.14$. We determined an
intrinsic scatter $\delta_{\rm bulge}=0.41\pm0.07$ which is $\sim$50\%
larger than the scatter in the $M_{\rm bh}-\sigma$ relation.

\end{abstract}


\section{Introduction}

Observations of bulges in nearby galaxies reveal the presence of
massive dark objects whose dynamical influence on the surrounding
stars is consistent with their being SMBHs (e.g. Kormandy \&
Richstone~1995). Moreover, the masses of these SMBHs correlate with
properties of the host galaxy, including the luminosity of the bulge
(Kormendy \& Richstone~1995), the mass of the bulge (Magorrian et
al.~1998), the stellar velocity dispersion of the bulge (Ferrarese \&
Merritt~2000; Gebhardt et al.~2000) and the concentration of the bulge
(Graham et al.~2002). The tightest relation, with intrinsic scatter of
$\sim0.3$ dex, appears to be between SMBH mass and bulge velocity dispersion
(Tremaine et al.~2002, hereafter T02). Significant effort has been
invested in determining parameters that describe this relation, which
is usually parameterised using the log-linear functional form
\begin{equation}
\label{fit}
\log(M_{\rm bh})=\alpha + \beta\log(\sigma/200{\rm km }\,{\rm s}^{-1}).
\end{equation} 
The value of the power-law slope $\beta$ has been a matter of some
debate (see T02 for a review). Recent estimates have
$\beta=4.02\pm0.33$ (T02) and $\beta=4.83\pm0.43$
(Ferrarese \& Ford~2004). These values differ by $\sim$2-sigma, a difference
which may be attributable to systematic differences in the velocity
dispersions used by different groups (T02).

Although the connection between SMBHs and their host galaxies is not
yet clear, it seems very likely that their evolution is closely
intertwined. It also seems likely, given the small intrinsic scatter
in the relation that the value of $\beta$ will yield important clues
regarding the physics of SMBH evolution (e.g. Silk \& Rees~1998;
Wyithe \& Loeb~2003; King~2004; Miralda-Escude~2004; Saznov et
al.~2005). Moreover, studies of SMBH demographics (e.g. Yu \&
Tremaine~2002; Shankar et al.~2004) rely on this relation to estimate
quantities of astrophysical interest like the local SMBH mass density. For
these reasons, and given the heroic efforts that have been made to
measure SMBH masses and host velocity dispersions, it is important to
make statistically robust estimates of parameters that describe the
$M_{\rm bh}-\sigma$ relation.

Rather than assume a powerlaw relation between $M_{\rm bh}$ and
$\sigma$ we take a more general approach in this paper. Suppose we have an unknown
relation $\log(M_{\rm bh})=f(\log(\sigma))$. This relation may be
expanded in a Taylor series in $\log{\sigma}$ about $\sigma=200$km/s,
which yields to second order
\begin{eqnarray}
\label{fit_quad0}
\nonumber
\log(M_{\rm bh})=\alpha &+& \beta\log(\sigma/200{\rm km }\,{\rm s}^{-1})\\
 &+& \beta_2\left[\log(\sigma/200{\rm km }\,{\rm s}^{-1})\right]^2.
\end{eqnarray} 
The zeroth and first order coefficients in this expansion correspond
to $\alpha$ and $\beta$ in the usual log-linear relation.  However in addition
to $\alpha$ and $\beta$ we also consider the possibility of a log-quadratic
term. The parameter $\beta_2$ therefore provides a measure of the
deviation from a pure powerlaw of the $M_{\rm bh}-\sigma$ relation.

We show that there is a non-linear contribution to the
$\log{M_{\rm bh}}-\log{\sigma}$ relation, and that the assumption of a
log-linear relation has biased estimates of the power-law
slope. For the Local sample of T02 we show that the variation in log-linear
slope over the range of $\sigma$ in the sample is larger than both the
statistical uncertainty in the slope of the log-linear relation and
the 2-sigma difference between the estimates of T02 and
Ferrarese~(2002). Note that (with the exceptions of \S~\ref{choice} and the appendices) we
restrict our attention to the statistical bias in estimation of the
parameters $\beta$ and $\beta_2$ from SMBH masses and galaxy velocity
dispersions summarised in T02. For further discussion of possible systematic
errors in the measured parameters $M_{\rm bh}$ and $\sigma$ themselves
we refer the reader to T02 and to Merritt \& Ferrarese~(2001).

An outline of the paper is as follows. We first discuss possible bias
in the local sample of SMBHs with respect to parameter fitting
(\S~\ref{biasfit}). Then in
\S~\ref{Msig} we repeat the parameter fitting analysis of T02 and
demonstrate that the value of $\beta$ in a log-linear relation depends
systematically on which low velocity dispersion galaxies are included
in the sample. We then show that a log-quadratic form for the $M_{\rm
bh}-\sigma$ relation provides an improved fit which is not
systematically sensitive to the galaxy sample (\S~\ref{quadMsig}).
The dependence of this conclusion on the method of SMBH mass
estimation and resolution of the SMBH sphere of influence, as well as
the definition of the velocity dispersion is discussed in
\S~\ref{choice}.  A Bayesian approach to parameter estimation in the
$M_{\rm bh}-\sigma$ relation and quantification of the systematic bias
in parameter estimation are described in \S~\ref{bayes}. This approach
allows an unbiased analysis for the Local sample which is presented in
\S~\ref{results}.  In \S~\ref{results_bulge} we apply our analysis to 
the relation between SMBH and bulge mass using the sample of Haering
\& Rix~(2004). We present some discussion in
\S~\ref{discussion}, before summarising our conclusions in \S~\ref{conclusion}.
Finally, in two appendices we analyse alternative samples to the one
described in T02.

\section{Parameter Estimation in The local SMBH sample}
\label{biasfit}

A first question that should be considered concerns whether or not the
local SMBH sample, and therefore the resulting estimates of the $M_{\rm
bh}-\sigma$ relation are statistically fair with respect to parameter
estimation.  The best-fit log-linear $M_{\rm bh}-\sigma$ relation
(T02; Ferrarese~2002) drops below $\sim2\times10^6M_\odot$ for
velocity dispersions smaller than $\sim70$km/s, still within the range
of $\sigma$ in the Local sample. However no kinematic detections of
SMBH masses have been published below $2\times10^6M_\odot$. Does this
lack of low mass SMBHs result in biased estimates of the slope in a
log-linear $M_{\rm bh}-\sigma$ relation?  Ferrarese~(2003) has
suggested that the reason for the dearth of SMBHs with $M<10^6M_\odot$
is partly instrumental sensitivity. SMBHs with masses smaller than
$10^6M_\odot$ are expected to reside in dwarf galaxies, with velocity
dispersions below 50km/s. The sphere of influence of these SMBHs would
be observable only in the most nearby cases. Ferrarese~(2003) has
produced a scatter plot of the local SMBH population in SMBH mass and
distance based on a combination of the CfA redshift survey (Huchra et
al.~1990) and the relation between SMBH mass and bulge luminosity. For
the nearest group of galaxies at $\sim3$Mpc, resolving the SMBH sphere
of influence limits detection to SMBHs with masses in excess of
$2\times10^6M_\odot$. Ferrarese~(2003) goes on to suggest that in the
nearby dwarf galaxies, individual stars become resolved, but are too
faint to allow dynamical studies, and that enlargement of the
kinematically detected sample to include SMBHs below $\sim10^6M_\odot$
will require campaigns using telescopes with apertures in excess of
8m, combined with resolution of better than 0.02''. One might
therefore suppose that there is a selection bias in the SMBH sample
for SMBHs that are larger than $\sim10^6M_\odot$. Such a selection
bias would lead to biased estimates of the $M_{\rm bh}-\sigma$
relation. However inspection of the Local sample shows that one of the
smallest SMBHs observed (N7457) lies at a distance well beyond where
its sphere of influence would have been resolved by HST
(Ferrarese~2003). Moreover, in the velocity range 70km$\,$s$^{-1}<\sigma<$380km$\,$s$^{-1}$ there are no galaxies in which a kinematic
search has been conducted by the Nuker team, and where a SMBH was not
found (S. Tremaine private communication). Since these galaxies are
not selected based on their expected SMBH mass, there should be no bias
against selection of low mass (or high mass) SMBHs {\em at a fixed
velocity dispersion}. One can therefore estimate the expected SMBH
mass given a velocity dispersion from the local SMBH sample, without
bias introduced through selection.

On the other hand, the sample {\em is} biased if one wants to estimate
the velocity dispersion at a fixed SMBH mass. The low surface
brightness of galaxies with large velocity dispersions results in a
maximum velocity dispersion where SMBHs can be measured (Ferrarese~2003). This combined
with the exponential decline in the number density of such galaxies
may result in an upper cutoff in the velocity dispersions found in the
observed sample. A parametric fit that treated velocity dispersion
rather than SMBH mass as the dependent variable would therefore lead
to a biased estimate of the parameters in the $M_{\rm bh}-\sigma$
relation.

\begin{figure*}
\vspace*{60mm}
\includegraphics{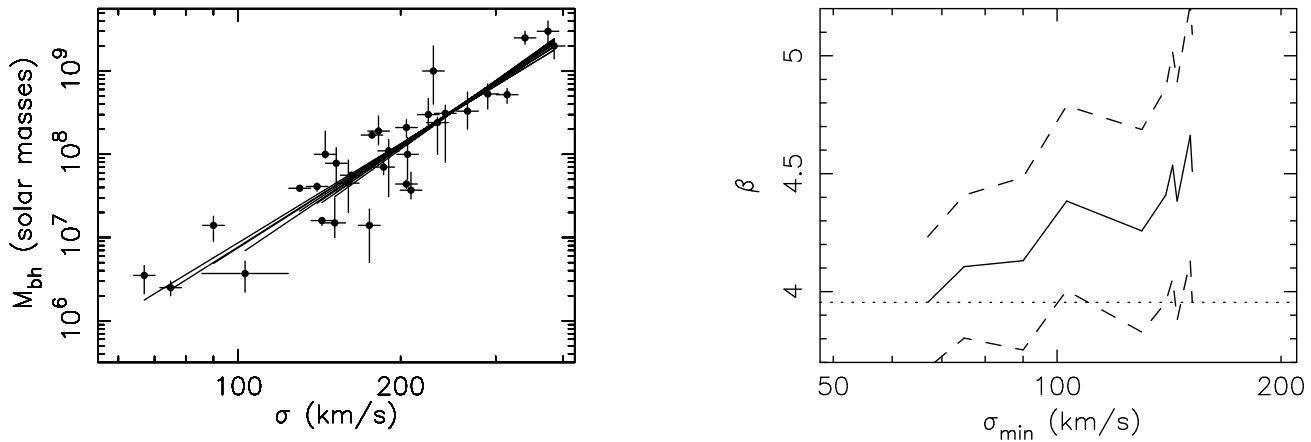}
\caption{\label{fig1} {\em Left:} The $M_{\rm bh}-\sigma$ relation from Tremaine et al.~(2002). {\em Right:} The 16th, 50th (solid line) and 84th percentiles of the probability distribution for $\beta$ obtained by minimising $\chi^2$ as a function of the minimum $\sigma$ considered in the sample. The best fit relations for these minimum dispersions are shown in the left panel, and demonstrate the effect of the lowest velocity points on the most likely fit.}
\end{figure*}

Clearly the selection function for the sample of SMBHs is complex. In
particular, the distribution of velocity dispersions in the sample
does not reflect the overall distribution of galaxies. The
non-uniformity of the distribution of velocity dispersions results in
statistical inferences for parameters in equation~(\ref{fit}) that
vary with the statistic used (Merritt \& Ferrarese~2001). This implies
that some or all of the estimates are biased. In order to make an
unbiased estimate for a parameter like the power-law slope $\beta$ one must
determine the bias inherent in the particular statistic used. In this
paper we overcome the selection bias introduced by the inhomogeneous
distribution of velocity dispersions via an investigation of the SMBH
mass vs. velocity dispersion relation in mock Monte-Carlo samples.

Given that we have access to a sample which is unbiased in $M_{\rm
bh}$ as a function of $\sigma$ (the Local sample, summarised in T02),
and in the absence of a theory where velocity dispersion is regulated
by SMBH mass (theoretical prejudice), we follow tradition and restrict
our attention to parameter fitting where $\log{M_{\rm bh}}$ is the
dependent variable as a function of $\log{\sigma}$.

\section{The Log-Linear $M_{\rm bh}-\sigma$ Relation}
\label{Msig}

Tremaine et al.~(2002) have compiled a list of $N_{\rm
g}=31$ galaxies (the Local sample) with reliable determinations of both SMBH mass
($M_{\rm bh}$) and central velocity dispersion ($\sigma$, defined as
the luminosity weighted dispersion in a slit aperture of half length
$R_{\rm e}$, the effective radius of the spheroid). The
sample\footnote{In this paper we refer to the Local sample as the sample of SMBHs and effective velocity dispersion (with uncertainties) listed in table~1 of T02. The exception is the Milky-Way galaxy for which we use the updated estimate for SMBH mass of $M_{\rm bh}=(3.7\pm1.5)\times10^6M_\odot$ (Sch{\" o}del et al.~2002). References for SMBH masses in table~1 of T02 are Verolme et al. (2002); Tremaine (1995); Kormendy \& Bender (1999); Bacon et al. (2001); Gebhardt et al. (2003); Bower et al. (2001); Greenhill \& Gwinn (1997); Sarzi et al. (2001); Kormendy et al. (1996); Barth et al. (2001); Kormendy et al. (1998); Gebhardt et al. (2000); Herrnstein et al. (1999); Ferrarese, Ford, \& Jaffe (1996); Cretton \& van den Bosch (1999); Harms et al. (1994); Macchetto et al. (1997); Ferrarese \& Ford (1999); van der Marel \& van den Bosch (1998); Cappellari et al. (2002). } is shown in the left panel of figure~\ref{fig1} which
illustrates the correlation between $M_{\rm bh}$ and $\sigma$.

We begin by repeating the analysis of T02 who estimate the parameters
$\alpha$ and $\beta$ through minimisation of a $\chi^2$ variable that
accounts for uncertainties in both $\sigma$ and $M_{\rm bh}$. The
$\chi^2$ variable used is
\begin{equation}
\label{chi2}
\chi^2 = \sum_{i=1}^{N_{\rm g}}\frac{(y_i-\alpha - \beta x_i)^2}{\epsilon_{yi}^2+\beta^2\epsilon_{xi}^2},
\end{equation}
where $y_i$ and $x_i$ are the logarithm of SMBH mass in solar masses
and the logarithm of velocity dispersion in units of 200km/s
respectively. The variables $\epsilon_{xi}$ and $\epsilon_{yi}$ are
the uncertainties in dex for these parameters. This expression
(equation~\ref{chi2}) is symmetric in $\log{M_{\rm bh}}$ and
$\log{\sigma}$. One might expect this to be a favourable property since
the fit does not include any preconceived notions of the physical
origin of the relation. In T02 an estimate of the intrinsic scatter
$(\epsilon_{\rm intrins})$ in the $M_{\rm bh}-\sigma$ relation [as
defined in the $y$ (or $M_{\rm bh}$) direction] was established by
adding $\epsilon_{\rm intrins}^2$ to the denominator in
equation~(\ref{chi2}). The intrinsic scatter that resulted in a
reduced minimum $\chi^2$ of unity was $\epsilon_{\rm intrins}\sim0.27$
dex. The best fit solution for a linear $M_{\rm bh}-\sigma$ relation
from T02 has $\alpha=8.13$ and $\beta=4.02$, resulting in residuals
for the three smallest galaxies that are greater than zero. In
addition, the largest three galaxies also have residuals that are
greater than zero. This behaviour is symptomatic of a scenario where a
linear relation has been fitted to a non-linear sample.

The sample described in T02 is dominated by galaxies with velocity
dispersions in the range $\sim130-250$km/s. There are a handful of
SMBHs at the centers of galaxies with smaller velocity dispersions,
including the Milky-Way. These SMBHs all have masses in excess of
$M_{\rm bh}\sim2\times10^6M_\odot$. The four galaxies with the lowest
dispersions have a large influence over the slope inferred for a
log-linear relation. Figure~\ref{fig1} shows how the estimate of
$\beta$ varies as the galaxies with the lowest velocity dispersions
are removed from the sample. Two trends are apparent. Firstly, the
uncertainty in $\beta$ is reduced by more than 50\% by the presence of
the smallest few galaxies. Secondly, the lowest velocity dispersions
reduce the estimate of the slope from $\beta\sim4.5$ to
$\beta\sim4$. The estimate remains near $\beta\sim4.5\pm0.5$ as the
next 6 smallest velocities are removed from the sample. The systematic
variation may be seen visually in the left panel of figure~\ref{fig1},
where the best-fit relations are plotted over the corresponding ranges
of velocity dispersion.

\begin{figure*}
\vspace*{130mm}
\includegraphics{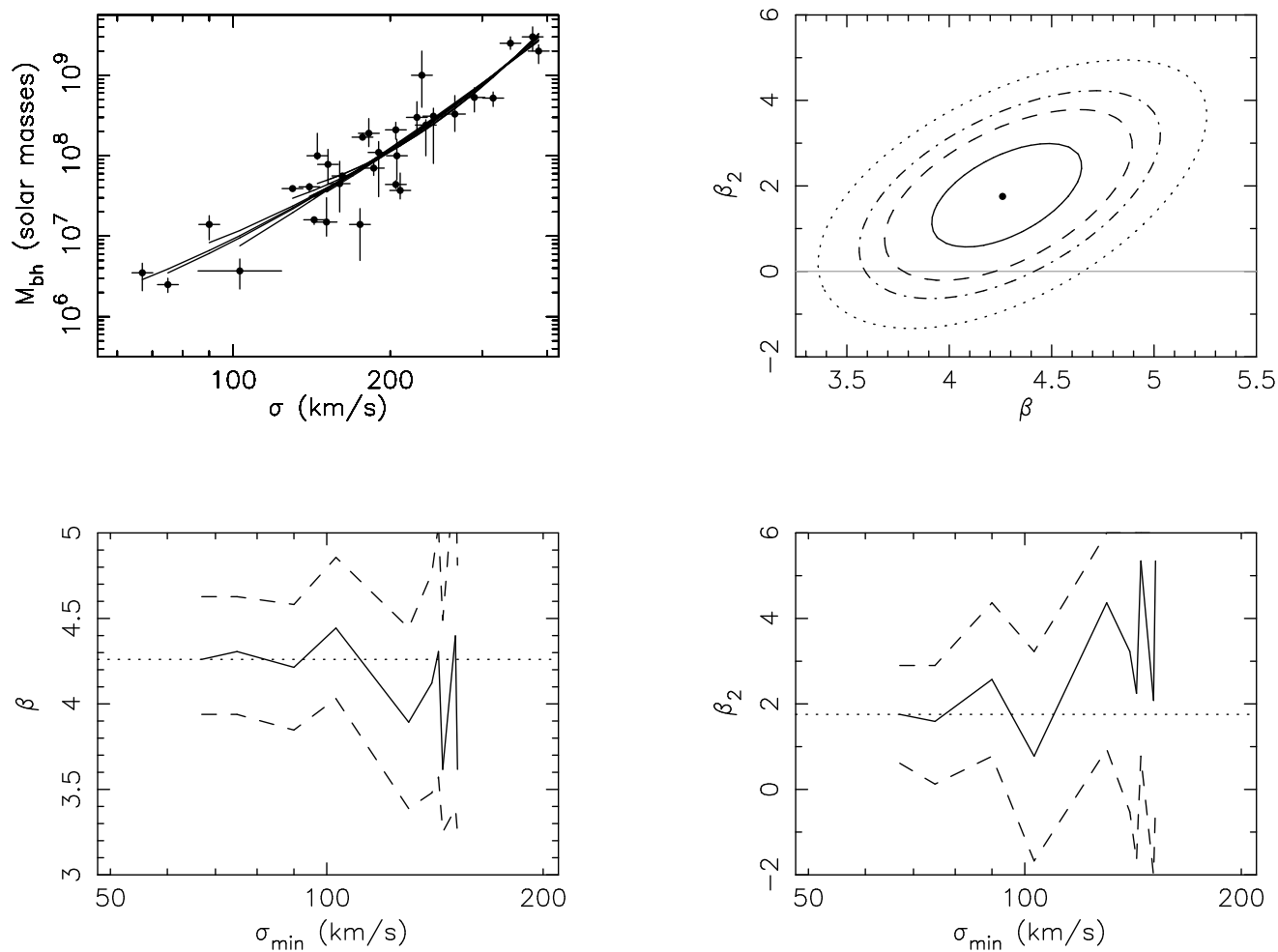}
\caption{\label{fig2} $\chi^2$ minimisation of the log-quadratic $M_{\rm bh}-\sigma$ relation. {\em Upper Right:} Contours of $\chi^2$ (minimised over $\alpha$) of $\beta$ and $\beta_2$. When projected, the contours correspond to the 1, 1.5, 2 and 2.5-sigma uncertainties on individual parameters. The point shows the most likely solution. {\em Lower Left:} The 16th, 50th (solid line) and 84th percentiles of the probability distribution for $\beta$ obtained by minimising $\chi^2$ as a function of the minimum $\sigma$ considered in the sample. {\em Lower Right:} The 16th, 50th (solid line) and 84th percentiles of the probability distribution for $\beta_2$ obtained by minimising $\chi^2$ as a function of the minimum $\sigma$ considered in the sample. The best fit relations for these minimum dispersions are shown in the {\em upper-left} panel, and demonstrate the effect of the lowest velocity points on the most likely fit.}
\end{figure*}

\section{A log-Quadratic $M_{\rm bh}-\sigma$ relation}
\label{quadMsig}

The systematic trend of a slope that increases by 1.5-sigma when the
smallest 3 of $N_{\rm g}=31$ galaxies are removed from the sample
points to a fit that is systematically biased. One possibility is that
a log-linear relation is not a good description of the data set. Note
that a poor fit cannot be identified through a reduced value of
$\chi^2$ that is significantly greater than unity. This is because the
value of intrinsic scatter ($\epsilon_{\rm intrins}$) is determined by
the condition that the reduced value of $\chi^2$ equal unity for the
best fit solution. However we can try to find a functional form that
provides a better description of the data. One possible scenario
involves a characteristic velocity dispersion, with different
powerlaws for large and small galaxies.  To test this idea we attempt
to locate a value of $\sigma$ around which the slope of the relation
changes by fitting two power-law slopes, one either side of a
characteristic velocity dispersion. We find that the fit is improved
by allowing the slope at high $\sigma$ to be steeper than at low
$\sigma$, and that the data prefers a break velocity somewhere in the
range $150-350$km/s (1-sigma). Of course, while there is at least some
theoretical motivation for a log-linear relation, there is no reason
to suppose that the $M_{\rm bh}-\sigma$ relation should be a double
power-law.

In the following we take a more general approach. As discussed in the
introduction, a general relation $\log{M_{\rm
bh}}=f(\log{\sigma})$ can be expanded in a
Taylor series to second order yielding
\begin{eqnarray}
\label{fit_quad}
\nonumber
\log(M_{\rm bh})&=&\alpha + \beta\log(\sigma/200{\rm km }\,{\rm s}^{-1})\\
 &+& \beta_2\left[\log(\sigma/200{\rm km }\,{\rm s}^{-1})\right]^2.
\end{eqnarray} 
Here the coefficients $\beta$ and $\beta_2$ represent log-linear and
log-quadratic contributions to the $M_{\rm bh}-\sigma$ relation. We
have (arbitrarily) expanded about $\sigma=200$km/s, which corresponds
to the median velocity dispersion in the relation. The value of
$\beta_2$ provides a measure of whether or not a log-linear relation
provides a good description of the data.

We have repeated the $\chi^2$ minimisation using
equation~(\ref{fit_quad}).  The 1, 1.5, 2 and 2.5-sigma ellipsoids\footnote{These ellipsoids correspond to loci where the difference between $\chi^2(\beta,\beta_2)$ and the minimum
$\chi^2$ is 1, 2.71, 4 and 6.63 respectively. When these contours are projected onto an individual parameter axis, they correspond to the 68\%, 90\%, 95\% and 99\% confidence intervals on those parameters. In the text we refer to these as the 1, 1.5, 2 and 2.5-sigma error elipsoids.} for
$\beta$ and $\beta_2$ are shown in the upper right-hand panel of
figure~\ref{fig2} for the Local sample. An intrinsic scatter of
$\epsilon_{\rm intrins}=0.25$ dex results in a reduced $\chi^2$ of
unity for the best fit solution. Thus a log-quadratic fit admits a
slightly smaller intrinsic scatter than the log-linear relation. The
solution shows evidence for a positive log-quadratic term at around the 85\% level. The most likely solution has $\beta=4.3$ and
$\beta_2=1.9$ with 1-sigma uncertainties of 0.35 and 1.1
respectively. The best fit log-quadratic relation is plotted in the
upper-left panel of figure~\ref{fig2}.

Parameter estimation using a log-linear relation corresponds to the
conditional probability for $\beta$ given $\beta_2=0$. The horizontal
grey line in the upper right panel of figure~\ref{fig2} shows the cut
through the bi-variate probability distribution for $\beta$ and
$\beta_2$. The contours of the bi-variate distribution cross this
line centered around $\beta=4$, which is consistent with expectations from the
log-linear fit. However the most probable value $\beta=4.02$ for the
log-linear fit lies near the 1.5-sigma contour of the log-quadratic
fit. This illustrates the point that a log-quadratic form provides a
much improved description of the Local sample.

Unlike the log-linear case, the four galaxies with the lowest
dispersions do not have a large influence on the log-quadratic
fit. Figure~\ref{fig2} shows that the estimates of $\beta$ and
$\beta_2$ do not vary systematically as the galaxies with the lowest
velocity dispersions are removed from the sample. The most likely
solution is similar for samples including all galaxies and for samples
including only the largest 20 galaxies, indicating that there is evidence
for a log-quadratic term in the main group of massive galaxies, and
that the deviation from a power-law is not dominated by inclusion of
low mass galaxies in the sample. The variation of the best-fit
relation as small galaxies are removed from the sample may be seen
visually in the left panel of figure~\ref{fig2}, where the best-fit
relations are plotted over the corresponding ranges of velocity
dispersion.

The above discussion relates to an expansion of the general relation
$\log{M_{\rm bh}}=f(\log{\sigma})$ which is truncated at second
order. We found that the data prefers a non-zero contribution from a
quadratic term. Before continuing we mention the possibility of a
non-zero contribution from an additional log-cubic term. We have
repeated the above analysis for the third order relation
\begin{eqnarray}
\label{fit_trip}
\nonumber
\log(M_{\rm bh})=\alpha &+& \beta\log(\sigma/200{\rm km }\,{\rm s}^{-1})\\
\nonumber
 &+& \beta_2\left[\log(\sigma/200{\rm km }\,{\rm s}^{-1})\right]^2\\
 &+& \beta_3\left[\log(\sigma/200{\rm km }\,{\rm s}^{-1})\right]^3.
\end{eqnarray} 
By minimising $\chi^2$ over the other parameters, we find a solution
$\beta=4.33\pm0.6$, $\beta_2=1.85\pm2.0$ and $\beta_3=0.3\pm6.0$. Not
surprisingly there is a large correlation between the odd terms
$\beta$ and $\beta_3$. This correlation increases the uncertainties on
individual parameters. Despite having an additional free parameter,
the log-cubic relation requires a larger intrinsic scatter to achieve
a $\chi^2$ of unity than the log-quadratic relation. Moreover there is
no evidence from the $\chi^2$ minimization for a log-cubic term in the
Local sample. Therefore in the remainder of the paper we restrict our
attention to the log-quadratic form for the $M_{\rm bh}-\sigma$
relation.

\section{Choice of SMBH sample and definition of $\sigma$}
\label{choice}

\begin{figure*}
\vspace*{130mm}
\includegraphics{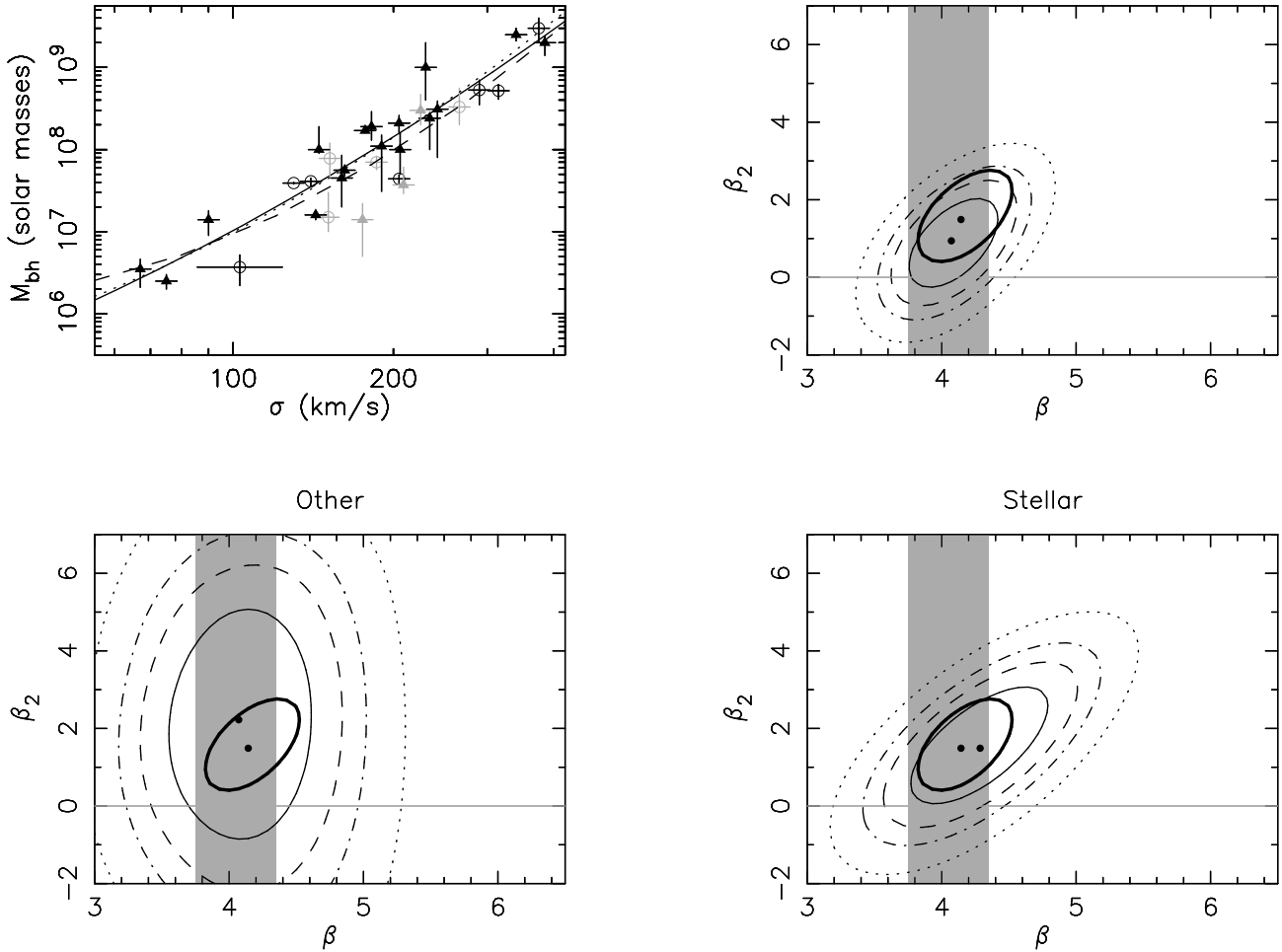}
\caption{\label{fig3} The upper left panel shows the Local sample with 
SMBH mass plotted vs effective velocity dispersion (T02).  The
triangles and open circles represent SMBH masses determined via
stellar dynamics and by other methods (masers, gas-disks, stellar
proper motions) respectively. The grey symbols denote SMBHs whose mass
estimates were deemed unreliable, or whose sphere of influence was not
resolved (Marconi \& Hunt~2003).  Log-quadratic fits were performed
on three samples. In the upper right panel we show a fit to the 24
galaxies whose spheres of influence are resolved, and whose masses are
deemed reliable. In the lower right panel we show a fit to the 20
galaxies with SMBH masses determined via stellar dynamics. In the
lower left panel we show a fit to 11 galaxies with masses determined
from methods other than stellar dynamics. Projections of the contours represent the
1, 1.5, 2 and 2.5-sigma uncertainties on individual parameters. The 1-sigma error ellipse (thick
line) for the full Local sample is shown superimposed on all three
panels. The grey region shows the corresponding linear fit value for
$\beta$ from T02. In the upper left panel we show the three best
fits. }
\end{figure*}

\begin{figure*}
\vspace*{130mm}
\includegraphics{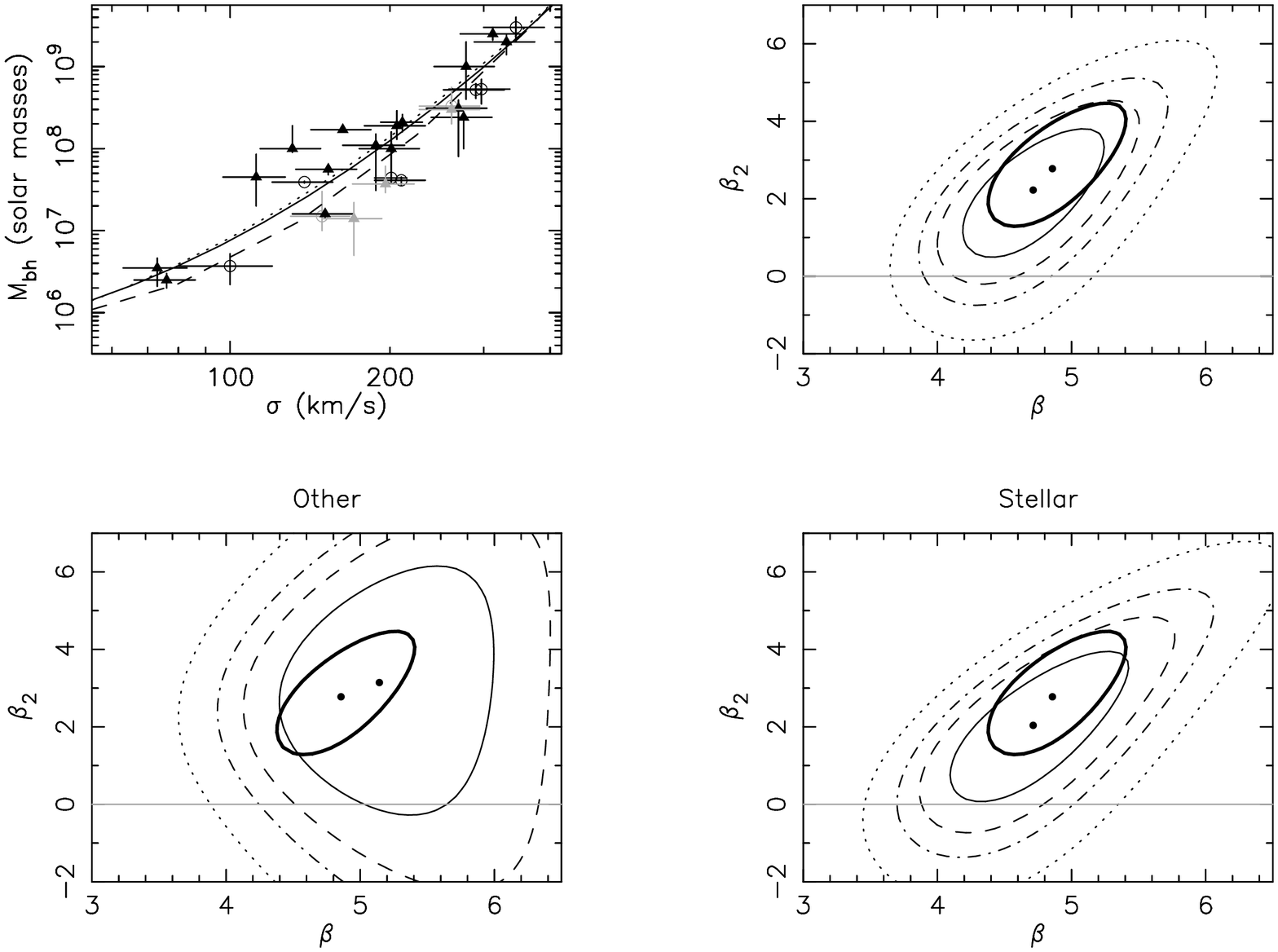}
\caption{\label{fig4} The upper left panel shows the Local sample
with SMBH mass plotted vs central velocity dispersion (Ferrarese \&
Merritt~2000). The triangles and open circles represent SMBH masses
determined via stellar dynamics and by other methods (masers,
gas-disks, stellar proper motions) respectively. The grey symbols
denote SMBHs whose mass estimates were deemed unreliable, or whose
sphere of influence was not resolved (Marconi \& Hunt~2003).
Log-quadratic fits were performed on three samples. In the upper right
panel we show a fit to the 24 galaxies whose spheres of influence are
resolved, and whose masses are deemed reliable. In the lower right
panel we show a fit to the 20 galaxies with SMBH masses determined via
stellar dynamics. In the lower left panel we show a fit to 11
galaxies with masses determined from methods other than stellar
dynamics. Projections of the contours represent the 1, 1.5, 2 and 2.5-sigma uncertainties on individual parameters. The 1-sigma error ellipse (thick line) for the full Local
sample is shown superimposed on all three panels. In the
upper left panel we show the three best fits. }
\end{figure*}

The Local SMBH sample described in T02 contains galaxies with SMBH mass
estimates obtained from data that did not resolve the sphere of
influence (e.g. Marconi \& Hunt~2003), and the accuracy of these masses has
been called into question (e.g. Merritt \& Ferrarese~2001; Ferrarese
\& Ford~2004; Marconi \& Hunt~2003). In addition, the sample has SMBHs 
with masses inferred using four separate techniques, including stellar
dynamics, stellar proper motions, astrophysical masers and dynamics of
gaseous disks. Moreover, adding to this inhomogeneity are the
different definitions for the velocity dispersion that have been supported
by different groups. In this section we investigate the effect that
these properties of the sample may have on the conclusion of a
log-quadratic $M_{\rm bh}-\sigma$ relation. These issues are also
discussed further in two appendices (\S~\ref{app1} and \ref{app2}).

Some of the original SMBH mass estimates based on ground based stellar
dynamical data (Magorrian et al.~1998) lie systematically above the
$M_{\rm bh}-\sigma$ relation defined by other galaxies in the local
sample with {\it more secure} mass estimates by as much as 2 orders of
magnitude (Ferrarese \& Merritt~2000). The correlation between the
offset and the galaxies distance has been interpreted as evidence that
the lack of resolution in these early observations resulted in a
systematic resolution dependent error in the Magorrian et al.~(1998)
masses (Merritt \& Ferrarese~2001). Of course it is not clear that one
can assume masses to be incorrect because they don't agree with the
log-linear $M_{\rm bh}-\sigma$ relation as described by the remaining
SMBHs. This would require the argument that the {\it observational}
mass estimates must be wrong because they disagree with the a {\it
theoretical} preconception that the $M_{\rm bh}-\sigma$ relation is
log-linear. However concern has been expressed that that SMBH masses
estimated using data that does not resolve the sphere of influence are
unreliable (e.g. Ferrarese \& Ford~2004). Marconi \& Hunt~(2003) have
ordered the SMBH sample in terms of the degree to which the SMBH
sphere of influence is resolved. There are 4 galaxies for which the
sphere of influence is smaller than the resolution (the
half-width-at-half-maximum) of the data. In addition, Marconi \&
Hunt~(2003) classed some of the mass estimates as unreliable for other
reasons such as unknown disk inclination, resulting in the removal of
a further 3 galaxies from the sample. In the upper left panel of
Figure~\ref{fig3} we show the remaining 24 galaxies (dark
points). Those SMBH masses designated by Marconi \& Hunt~(2003) as
being unreliable, and those with spheres of influence that are not
resolved are plotted in grey. In the right panel of Figure~\ref{fig3}
we show the log-quadratic fit to these 24 galaxies. The result should
be compared with the fit to the full sample, the 1-sigma error ellipse
for which is also plotted (thick contour). We see that the removal of
the 7 {\it unreliable} galaxies has slightly increased the allowed
range of $\beta_2$. However the solution still prefers a non-linear
term, the 1-sigma error ellipse lies near the $\beta_2=0$ line.

In the lower right- and lower-left panels of Figure~\ref{fig3} we
show the log-quadratic fits to the sample of galaxies with SMBH masses
determined via stellar dynamics, and by other methods
respectively. Each of these samples also shows evidence for a
non-linear contribution to the $M_{\rm bh}-\sigma$ relation. The sample containing SMBHs with resolved spheres of
influence, and the sample containing only SMBHs with masses determined
via stellar dynamics have very similar solutions, indicating that
SMBHs with masses determined via stellar dynamics in cases where the
sphere of influence is unresolved do not skew the analysis and may be
included in the sample. 

The question of the dependence of resolution on the accuracy of the
SMBH mass determined was investigated by Gebhardt et al.~(2003). They
re-evaluated the SMBH masses for galaxies where the SMBH mass had been
previously determined with ground-based imaging using higher
resolution space-based data. Gebhardt et al.~(2003) showed that {\it
i)} the Magorrian et al.~(1998) masses were indeed overestimated, by
up to a factor of 3, due to the assumption of 2-integral models; and
{\it ii)} comparison of 3-integral modeling of low-resolution ground
based and high resolution space based data demonstrates that the SMBH
mass determinations described in Gebhardt et al.~(2003) contain no
resolution induced bias. In other words, provided that the modeling is
sufficiently general, the high resolution data increases the precision
but not the accuracy of SMBH mass estimates (see figure~8 of Gebhardt
et al.~2003). This implies that if the resolution were not sufficient,
then the lower limit on BH mass would be $<0$ (i.e. no definite
detection). Thus the effect of spatial resolution is built into the
estimate of the range of allowable mass. The masses of SMBHs in
galaxies where the sphere of influence is not resolved should
therefore be just as reliable as those where it is resolved, in the
sense that the determination is just as accurate but carries less
precision (Richstone et al.~2004).

At this point it should be noted that recent work from Valluri,
Merritt \& Emsellem~(2004) has suggested that stellar dynamical
estimates of SMBH masses are unreliable even when the SMBH sphere of
influence is resolved. This unreliability is due to a degeneracy
between the mass-to-light ratio and SMBH mass, which was shown to
become more prominent as the number of orbits used in the modeling is
increased. On the other hand, Richstone et al.~(2004) have countered
this claim. They suggest that while it is true that the allowed range
of SMBH mass increases with the number of orbits assumed, this range
asymptotes to provide a true estimate of SMBH mass. Thus they
conclude that provided the number of orbits is sufficient, SMBH mass
estimation from reconstruction of stellar orbits using kinematic data
can lead to a robust SMBH mass estimate. Richstone et al.~(2004)
estimate the number of orbits that are required to achieve convergence
in the mass estimate and conclude that published mass estimates have
been made using a sufficiently large orbit library and are therefore
reliable.

Finally, the definition of the independent variable may be an
important factor in the evaluation of any non-linear contribution to
the $M_{\rm bh}-\sigma$ relation. A central velocity dispersion has
been advocated by Ferrarese \& Merritt~(2001), while an effective
velocity dispersion was used by Gebhardt et al.~(2000). The pros and
cons of these different variables were summarised by Merritt \&
Ferrarese~(2001). The central velocity dispersion is more easily
measured than the effective velocity dispersion which requires both
surface photometry and spatially resolved spectroscopy over the
effective radius of the galaxy. As a result, central velocity
dispersions have been measured for many galaxies, while the effective
velocity dispersions require significant effort to obtain the required
data. On the other hand the effective velocity dispersion better
represents the velocity dispersion over the stellar spheroid, and may
therefore be expected to offer a better representation of the depth of
the spheroids gravitational potential well. Moreover T02 found that
the SMBH itself could effect the value of the central velocity by up
to 30\% in some cases. In order to evaluate the effect of changing the
independent variable from an effective to a central velocity
dispersion, we repeat the above analysis on sub-samples of the Local
sample of SMBHs (values of central velocity dispersion are available
in Merritt \& Ferrarese~2001 \& Ferrarese \& Ford~2004). The results
are shown in Figure~\ref{fig4}, correspond to, and are qualitatively
similar to to those in Figure~\ref{fig3}, where the effective
velocity dispersion was the independent variable. In particular, if
the SMBHs whose spheres of influence are not resolved are removed from
the sample, then the sample still prefers a non-linear
contribution. This non-linearity is at higher significance where the
central rather than effective velocity dispersion is used.  Similarly,
samples of SMBHs with masses determined via stellar dynamics and by
other methods each prefer a non-linear $M_{\rm bh}-\sigma$
relation. This analysis shows the previously known result that use of
the central velocity dispersion results in an estimate of the $M_{\rm
bh}-\sigma$ relation that is steeper than where the effective velocity
dispersion is used. 

In summary, samples where
the SMBH masses were estimated by stellar dynamics, samples where the
SMBH masses were determined via other methods, and samples where the
SMBH masses are {\it secure} (Marconi
\& Hunt~2003), each have a fit that prefers the inclusion of a 
log-quadratic term.
The sample of SMBH masses determined via other methods shows the
strongest non-linear tendency, but with the largest uncertainty. This
non-linear component of the $M_{\rm bh}-\sigma$ relation is present
whether the independent variable is considered to be the central
velocity dispersion (Ferrarese \& Merritt~2000) or the effective
velocity dispersion (T02). In light of these results, we consider only
the full Local sample as described in T02, with the effective velocity
dispersion as the independent variable, in the remainder of this
paper.

\section{Bayesian Parameter Estimation}
\label{bayes}

Our aim in this paper is to determine the parameters $\alpha$, $\beta$
and $\beta_2$, as well as the intrinsic scatter (designated $\delta$)
of SMBHs around equation~(\ref{fit_quad}). Our general approach adds
two features relative to the $\chi^2$ analysis described in the
previous sections.  First, our approach explicitly includes the
intrinsic scatter (labeled $\delta$) about the mean $M_{\rm
bh}-\sigma$ relation as a third free parameter, and we consistently
solve for all of $\alpha$, $\beta$, $\beta_2$ and $\delta$. The second
improvement is to include the asymmetric errors quoted for $M_{\rm
bh}$.

In the absence of errors in $M_{\rm bh}$ and $\sigma$ we may define
the contribution $LH_i$ of SMBH $i$ to the likelihood
$LH_{\alpha,\beta,\beta_2,\delta}$ for the set
$\{\alpha,\beta,\beta_2,\delta\}$.  However both $M_{\rm bh}$ and
$\sigma$ contain observational uncertainty.  Each of the $LH_i$ must
therefore be averaged over the uncertainty in both $M_{\rm bh}$ and
$\sigma$, hence
\begin{eqnarray}
\label{LH_i}
\nonumber
LH_i &\equiv& \int_{-\infty}^\infty d\log{\sigma} \frac{dP_i}{d\log{\sigma}}    \int_{-\infty}^\infty d\log{M_{\rm bh}}\\
&&\hspace{30mm} \frac{dP_i}{d\log{M_{\rm bh}}} LH(\sigma,M_{\rm bh}).
\end{eqnarray}
The uncertainty in $\sigma$ is described by
$dP_i/d\log{\sigma}= N(d\log{\sigma_i},0.02)$ for all galaxies except
the Milky-Way for which, following T02 we assume a larger uncertainty,
$dP_i/d\log{\sigma}= N(d\log{\sigma_i},0.08)$. Here the notation
$N(x|\bar{x},\delta x)$ refers to the value of a Gaussian distribution
with mean $\bar{x}$ and variance $\delta x$ at $x$. The observed
uncertainty in $M_{{\rm bh},i}$ is assumed to follow a distribution
\begin{eqnarray}
\nonumber
\frac{dP_i}{d\log{M_{\rm bh}}}&=&N(\log{M_{\rm
bh}}|\log{M_{{\rm bh},i}},\Delta \log{M^{\rm low}_{{\rm bh},i}})\\
\nonumber
&&\hspace{33mm}\mbox{if}\hspace{3mm}M_{{\rm bh}}<M_{{\rm bh},i}\\
\nonumber 
&=&N(\log{M_{\rm bh}}|\log{M_{{\rm bh},i}},\Delta \log{M^{\rm upp}_{{\rm bh},i}})\\
&&\hspace{33mm}\mbox{if}\hspace{3mm}M_{{\rm bh}}\geq M_{{\rm bh},i}
\end{eqnarray}
Here we have defined $\Delta \log{M^{\rm
upp}_{\rm bh,i}}$ and $\Delta
\log{M^{\rm low}_{\rm bh,i}}$ as the uncertainty (in dex) above and below
the observed value for SMBH $i$. The values of $\alpha$, $\beta$, $\beta_2$ and
$\delta$ may then be estimated by maximising the product of
likeli-hoods for the $N_{\rm g}$ residuals
\begin{equation}
\nonumber
LH_{\alpha,\beta,\beta_2,\delta} = \Pi_{i=0}^{N_{\rm g}} LH_{i}
\end{equation}
Note that we have not yet specified the definition of
$LH(\sigma,M_{\rm bh})$. For a finite sample size, the solution for
$\alpha$, $\beta$, $\beta_2$ and $\delta$ is sensitive to the definition of
$LH(\sigma,M_{\rm bh})$. As a result, an inappropriate choice for
$LH(\sigma,M_{\rm bh})$ can lead to a biased estimate of the
parameters. In this paper we are analysing a sample where SMBH mass is
unbiased at a fixed value of $\sigma$. We therefore treat $\sigma$ as
the independent variable and model the intrinsic scatter as a Gaussian
with variance $\delta$, hence
\begin{eqnarray}
\label{LH1}
\nonumber
LH(\sigma,M_{\rm bh}) &\equiv& N(\Delta_{M_{\rm bh},i}  |0,\delta)\\
\nonumber
&&\hspace{-28mm}\mbox{where}\hspace{5mm}\Delta_{M_{\rm bh},i} = \log{M_{{\rm bh},i}}-\left[\alpha+\beta\log{\left(\frac{\sigma_i}{200\mbox{km/s}}\right)}\right.\\
&&\hspace{+20mm}\left.+\beta_2\log{\left(\frac{\sigma_i}{200\mbox{km/s}}\right)^2}\right]. 
\end{eqnarray}

\begin{figure*}
\vspace*{130mm}
\includegraphics{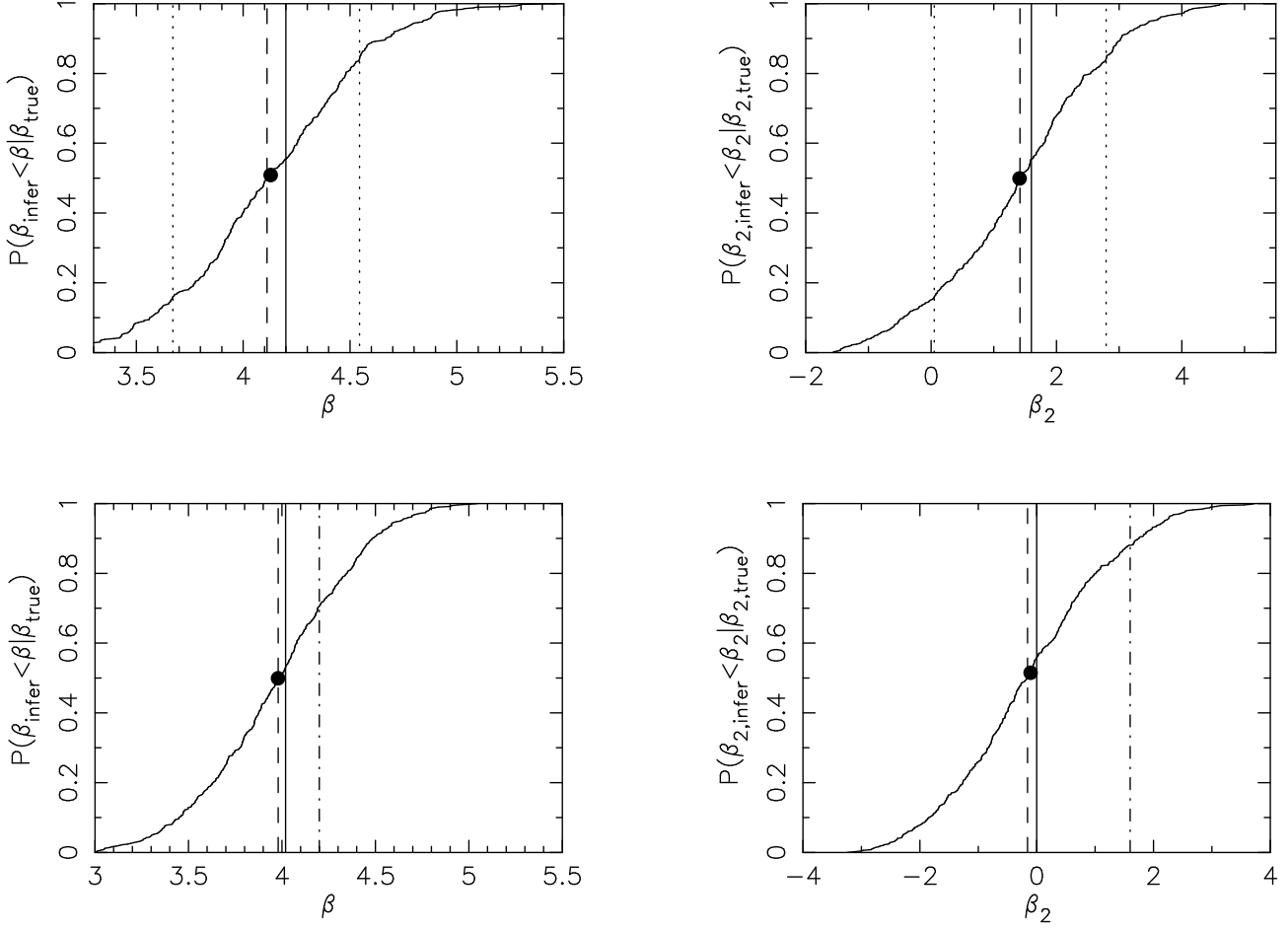}
\caption{\label{fig5} The cumulative distribution of maximum likelihood solutions for the inferred values of $\beta_{\rm infer}$ and $\beta_{\rm 2,infer}$  in Monte-Carlo samples given different input values $\beta_{\rm true}$ and $\beta_{\rm 2,true}$, which are shown by the vertical solid lines. The dots and dashed lines show the distribution means and medians respectively. {\em Upper panels:} The input relation had $\beta_{\rm true}=4.2$ and $\beta_{\rm 2,true}=1.6$. The dashed lines show the 16th and 84th percentiles. {\em Lower panels:} The input relation had $\beta_{\rm true}=4.02$ and $\beta_{\rm 2,true}=0.$ The dot-dashed lines show the values of $\beta$ and $\beta_2$ inferred from the data.}
\end{figure*}

The joint a-posteriori probability distributions for the parameters
$\alpha$, $\beta$ , $\beta_2$ and $\delta$ may be found from
\begin{equation}
\label{d4P}
\frac{d^4P}{d\alpha d\beta d\beta_2 d\delta}\propto LH_{\alpha,\beta,\beta_2,\delta}\frac{dP_{\rm prior}}{d\alpha}\frac{dP_{\rm prior}}{d\beta}\frac{dP_{\rm prior}}{d\beta_2}\frac{dP_{\rm prior}}{d\delta},
\end{equation}
where the $P_{\rm prior}$ are prior probabilities, which we assume to
be flat in this paper, i.e. 
\begin{equation}
\frac{dP_{\rm prior}}{d\alpha}\propto\frac{dP_{\rm prior}}{d\beta}\propto\frac{dP_{\rm prior}}{d\beta_2}\propto\frac{dP_{\rm prior}}{d\delta}\propto1.
\end{equation}
A-posteriori distributions for combinations of
these parameters may then be obtained by marginalising over the other
dimensions, for example
\begin{equation}
\label{marginalised}
\frac{d^2P}{d\delta d\beta}\propto\int_{-\infty}^\infty d\alpha \int_{-\infty}^\infty d\beta_2  \frac{d^4P}{d\alpha d\beta d\beta_2 d\delta}
\end{equation}
or
\begin{equation}
\label{marginalised2}
\frac{dP}{d\beta}\propto\int_{-\infty}^\infty d\alpha \int_{-\infty}^\infty d\beta_2 \int_{0}^\infty d\delta \frac{d^4P}{d\alpha d\beta d\beta_2 d\delta}.
\end{equation}

\subsection{Bias in Parameter Estimation}
\label{bias}

Before presenting the probability distributions for the parameters
$\beta$, $\beta_2$ and $\delta$ that describe the Local sample, we
asses the bias in the fitting procedure by fitting parameters to
Monte-Carlo realisations of mock SMBH samples. We generate samples of
$N_{\rm g}=31$ SMBHs using the following procedure. For each of the
$N_{\rm g}=31$ galaxies we select a value of $\sigma$ drawn from the
observed estimate $N(\sigma_i,\Delta\sigma_i)$. This value is used to
select a SMBH mass from an input mean relation, offset randomly
according to the input intrinsic scatter $\delta_{\rm true}$. This mass is then
further offset by a value drawn randomly from the quoted uncertainty
in $M_{\rm bh}$ for the corresponding SMBH in the observed sample. We
assumed input values of $\beta_{\rm true}=4.2$ and $\beta_{\rm
2,true}=1.6$. In the following section we show that these values lie
near the best fit relation for the Local sample. The intrinsic scatter
of the relations was assumed to be $\delta_{\rm true}=0.3$, defined in the $y$ or
$M_{\rm bh}$ direction.

For each mock sample we find the most likely values for $\beta$ and
$\beta_2$ from equation~(\ref{d4P}). The cumulative
distributions\footnote{Note that for a fair comparison, the most
likely velocities from the sample are used in the fitting procedure,
rather than the values of $\sigma$ used to calculate the mock SMBH
mass.} of the best fit solutions are plotted in the upper panels of
figure~\ref{fig5}. The estimates are not skewed, but the parameters
determined are slightly biased (by $\delta\beta=0.1$ and
$\delta\beta_2=0.2$) as may be seen by the comparing the median (dashed line)
and mean (denoted by the large dot) to the true value (solid vertical
line) in each case.

\section{Unbiased A-posteriori probability distributions for parameters in the log-quadratic $M_{\rm bh}-\sigma$ Relation}
\label{results}

\begin{figure*}
\vspace*{205mm}
\includegraphics{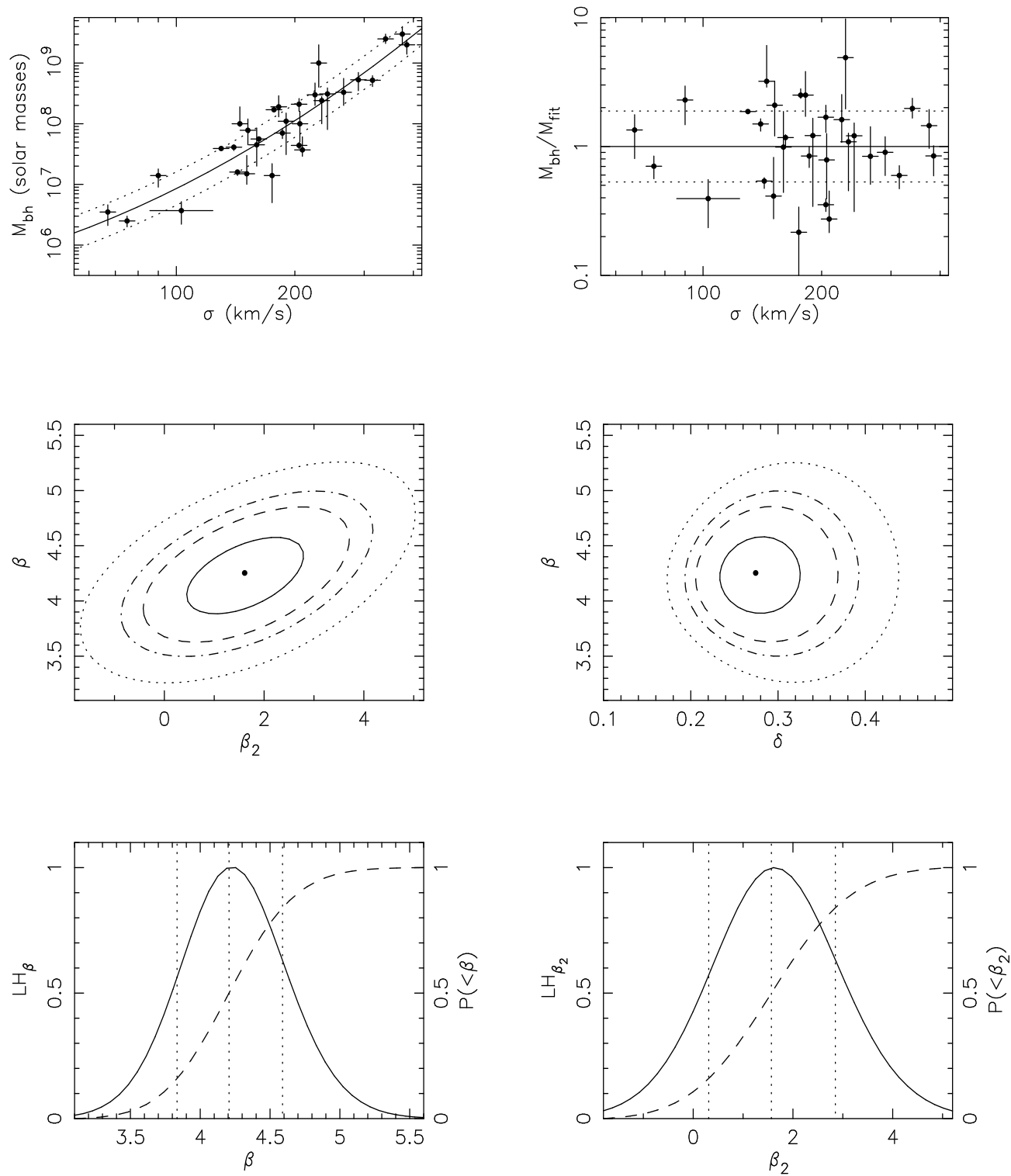}
\caption{\label{fig6} {\em Upper Left:} SMBH mass vs. central velocity dispersion from the compilation of T02. The solid line shows the best fit solution to equation~(\ref{fit_quad}) obtained using a likelihood defined in equation~(\ref{LH1}). The dotted lines show the level of intrinsic scatter around the best-fit relation. {\em Upper Right:} Residuals relative to the maximum likeli-hood solution. The dotted lines show the best fit value of intrinsic scatter. The lower panels show the a-posteriori probability distributions for the parameters $\beta$, $\beta_2$ and $\delta$. {\em Center Left:} Joint distribution for $\beta$ and $\beta_2$ marginalised over $\alpha$ and $\delta$. {\em Center Right:} Joint distribution for $\delta$ and $\beta$ marginalised over $\alpha$ and $\beta_2$. The contours refer to 0.036, 0.14, 0.26 and 0.64 of the peak height corresponding to the 4, 3, 2, and 1-sigma limits of a Gaussian distribution. {\em Bottom Left:} Differential (solid line) and cumulative (dashed line) distributions for $\beta$ marginalised over $\beta_2$, $\delta$ and $\alpha$. {\em Bottom Right:} Differential (solid line) and cumulative (dashed line) distributions for $\beta_2$ marginalised over $\beta$, $\delta$  and $\alpha$. The vertical dotted lines show the 16th, 50th and 84th percentiles. }
\end{figure*}

Having assessed the bias in our parameter estimation, we are in a
position to estimate the parameters describing the Local sample. In
figure~\ref{fig6} we show the a-posteriori marginalised probability
distributions for the parameters $\beta$, $\beta_2$ and $\delta$
computed using equations~(\ref{marginalised}-\ref{marginalised2}) in
combination with the likelihood equation~(\ref{LH1}). The values of
$\beta$ and $\beta_2$ corresponding to a fixed likelihood have been
corrected a-posteriori for the biases $\Delta\beta=0.1$ and
$\Delta\beta_2=0.2$. In the central row we show joint distributions
for $\beta_2$ and $\beta$ (left) and for $\delta$ and $\beta$
(right). The contours (dark lines) refer to 0.036, 0.14, 0.26 and 0.64
of the peak height corresponding to the 4, 3, 2, and 1-sigma limits of
a Gaussian distribution. In the bottom rows (dark lines) we show
differential (solid lines) and cumulative (dashed lines) distributions
for $\beta$ (left) and $\beta_2$ (right). The vertical dotted lines
show the variance.

The contours show little correlation between parameters. Using the
likelihood defined in equation~(\ref{LH1}) we find that the
marginalised distributions imply $\beta=4.2\pm0.37$ and
$\beta_2=1.6\pm1.3$ (figure~\ref{fig6}), and an intrinsic scatter of
$\delta=0.28\pm0.04$. The normalisation was found to be
$\alpha=8.05\pm0.06$. The best fit (solid line) is plotted over the
data in the upper left panel of figure~\ref{fig6}, together with
curves showing the level of intrinsic scatter (dotted lines). In the
upper right panel of figure~\ref{fig6} we plot the residuals in
$\log{M_{\rm bh}}$, together with horizontal dotted lines showing the
value of the best fit intrinsic scatter. We have assumed an intrinsic
scatter that is constant with $\sigma$. Inspection of the residuals in
figure~\ref{fig6} indicates that there is no systematic trend.

The cumulative probability distribution for $\beta_2$ shows that there
is a positive contribution (at $90\%$ confidence) from a log-quadratic
term in the $M_{\rm bh}-\sigma$ relation that describes the Local
sample. This may be interpreted as indicating that at the 80\% level
the $M_{\rm bh}-\sigma$ relation does not follow a single powerlaw
between $\sim70$km/s and $\sim380$km/s. An alternative statistical
question regarding the significance of this result concerns the
frequency with which one might measure a log-quadratic term as large
as the best fit of $\beta_2=1.6$, assuming an intrinsically log-linear
relation. We have performed fits to mock samples assuming an input
intrinsic slope of $\beta_{\rm true}=4.02$ and an input $\beta_{\rm
2,true}=0$, which corresponds to the best fit log-linear relation
(T02). The cumulative distributions for the best fit $\beta$ and
$\beta_2$ are plotted in the lower panels of figure~\ref{fig5}. We
find that the estimate of $\beta$ is unbiased in this case, and that
the best fit value of $\beta_2$ is greater than 1.6 around $10\%$ of the
time. This is consistent with the statement that the Local sample does
not follow a single power-law at the $\sim80\%$ level.

\begin{figure*}
\vspace*{205mm}
\includegraphics{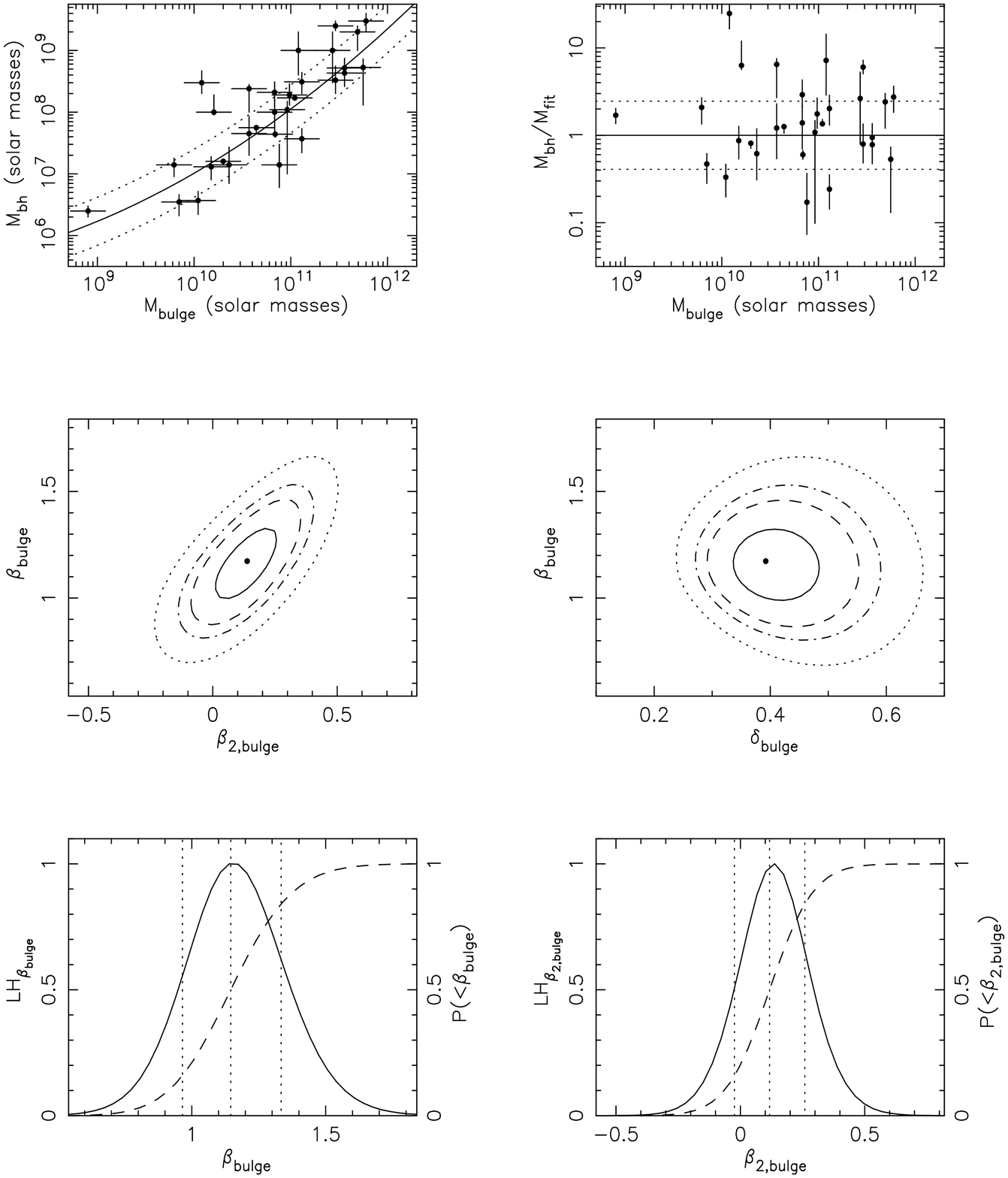}
\caption{\label{fig7} {\em Upper Left:} SMBH mass vs. bulge mass from the compilation of Haering \& Rix~(2004). The solid line shows the best fit solution to equation~(\ref{fit_quad_bulge}) obtained using a likelihood defined in equation~(\ref{LH1_bulge}). The dotted lines show the level of intrinsic scatter around the best-fit relation. {\em Upper Right:} Residuals relative to the maximum likeli-hood solution. The dotted lines show the best fit value of intrinsic scatter. The lower panels show the a-posteriori probability distributions for the parameters $\beta_{\rm bulge}$, $\beta_{\rm 2,bulge}$ and $\delta_{\rm bulge}$. {\em Center Left:} Joint distribution for $\beta_{\rm bulge}$ and $\beta_{\rm 2,bulge}$ marginalised over $\alpha_{\rm bulge}$ and $\delta_{\rm bulge}$. {\em Center Right:} Joint distribution for $\delta_{\rm bulge}$ and $\beta_{\rm bulge}$ marginalised over $\alpha_{\rm bulge}$ and $\beta_{\rm 2,bulge}$. The contours refer to 0.036, 0.14, 0.26 and 0.64 of the peak height corresponding to the 4, 3, 2, and 1-sigma limits of a Gaussian distribution. {\em Bottom Left:} Differential (solid line) and cumulative (dashed line) distributions for $\beta_{\rm bulge}$ marginalised over $\beta_{\rm 2,bulge}$, $\delta_{\rm bulge}$ and $\alpha_{\rm bulge}$. {\em Bottom Right:} Differential (solid line) and cumulative (dashed line) distributions for $\beta_{\rm 2,bulge}$ marginalised over $\beta$, $\delta$  and $\alpha$. The vertical dotted lines show the 16th, 50th and 84th percentiles. }
\end{figure*}

The inclusion of a log-quadratic term in the fit leads to residuals
that do not lie systematically above the mean relation at high or low
$\sigma$. We can therefore use the behaviour of these residuals to
interpret the significance of the log-quadratic term. Our Monte-Carlo
simulations (figure~\ref{fig5}) show that (after accounting for the
bias) $\sim10\%$ of samples with $\beta=4.2$ and $\beta_2=1.6$ produce
best fits with $\beta_2<0$.  As we saw in figure~\ref{fig1} the
smallest three galaxies all have SMBHs that are more massive than the
average implied by a log-linear fit. Each SMBH has $\sim1$ chance in 2
of lying above the mean relation. This situation would therefore arise
by chance for $\sim1$ sample in 10 ($\sim1/2^3$), which provides an
intuitive understanding of the finding that there is a log-quadratic
term in the $M_{\rm bh}-\sigma$ relation described by the Local sample
that is positive at the 90\% level.

\begin{figure*}
\vspace*{60mm}
\includegraphics{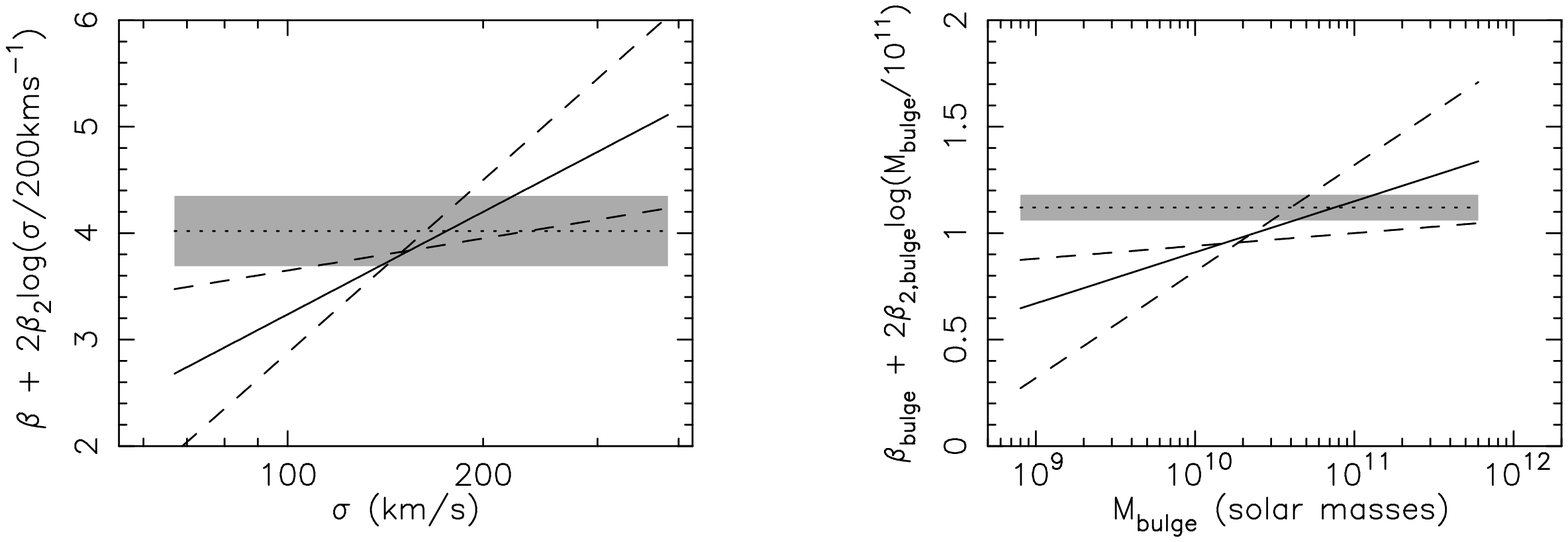}
\caption{\label{fig8} The values of logarithmic slope for the mean relations (solid lines) and for the relations using parameters on the edge of the 1-sigma error ellipses that lead to the most extreme examples (dashed lines). The logarithmic slopes are shown as a function of $\sigma$ for the $M_{\rm bh}-\sigma$ (left panel) and $M_{\rm bh}-M_{\rm bulge}$ relations (right panel). The grey regions represent the 1-sigma ranges for $\beta$ (T02) and $\beta_{\rm bulge}$ (Haering \& Rix~2004) using a log-linear relation. }
\end{figure*}

\section{The log-quadratic $M_{\rm bh}-M_{\rm bulge}$ Relation}
\label{results_bulge}

In this section we apply our analysis to the sample of SMBHs and bulge
masses summarised in Haering \& Rix~(2004). This sample closely
resembles that described in T02. The
SMBH-bulge mass relation is shown in the upper left panel of
figure~\ref{fig7}, and can be described by a log-quadratic relation of
the form
\begin{eqnarray}
\label{fit_quad_bulge}
\nonumber
\log(M_{\rm bh})=\alpha_{\rm bulge} &+& \beta_{\rm bulge}\log(M_{\rm bulge}/10^{11}M_\odot)\\
 &+& \beta_{\rm 2,bulge}\left[\log(M_{\rm bh}/10^{11}M_\odot)\right]^2.
\end{eqnarray} 
We have repeated our $\chi^2$ analysis on the $M_{\rm bh}-M_{\rm
bulge}$ sample (not shown). We find that the slope estimated using a
log-linear fit systematically increases by an amount that is greater
than the statistical uncertainty when the smallest galaxies are removed
from the sample. A $\chi^2$ minimisation of a log-quadratic relation
provides a significantly better fit, with the power-law relation ruled
out at the 1-sigma level.  The parameters of the log-quadratic fit do
not vary systematically as galaxies are removed from the sample.

Motivated by the success of the log-quadratic fit for the $M_{\rm
bh}-M_{\rm bulge}$ relation in a $\chi^2$ minimisation, we perform a
Bayesian analysis in analogy to the one described in the previous
section. We define a likeli-hood function
\begin{eqnarray}
\label{LH_i_bulge}
\nonumber
LH_i &\equiv& \int_{-\infty}^\infty d\log{M_{\rm bulge}} \frac{dP_i}{d\log{M_{\rm bulge}}}    \int_{-\infty}^\infty d\log{M_{\rm bh}}\\
&&\hspace{20mm} \frac{dP_i}{d\log{M_{\rm bh}}} LH(M_{\rm bulge},M_{\rm bh}),
\end{eqnarray}
where $LH(M_{\rm bulge},M_{\rm bh})$ is the likelihood of a set of
$M_{\rm bulge}$ and $M_{\rm bh}$ given a relation described by
$\alpha_{\rm bulge}$, $\beta_{\rm bulge}$, $\beta_{\rm 2,bulge}$ and
$\delta_{\rm bulge}$. Following Haering \& Rix~(2004) the uncertainty
in $M_{\rm bulge}$ is described by $dP_i/d\log{M_{\rm bulge}}=
N(d\log{M_{\rm bulge}},0.18)$ for all galaxies. At fixed bulge mass,
we model the intrinsic scatter as a Gaussian with variance
$\delta_{\rm bulge}$, hence we define a likeli-hood in analogy with
equation~(\ref{LH1})
\begin{eqnarray}
\label{LH1_bulge}
\nonumber
LH(M_{\rm bulge},M_{\rm bh}) &\equiv& N(\Delta_{M_{\rm bh},i}  |0,\delta_{\rm bulge})\\
\nonumber
&&\hspace{-31mm}\mbox{where}\hspace{3mm}\Delta_{M_{\rm bh},i} = \log{M_{{\rm bh},i}}\\
\nonumber
&&\hspace{-13mm}-\left[\alpha_{\rm bulge}+\beta_{\rm bulge}\log{\left(\frac{M_{{\rm bulge},i}}{5\times10^{10}M_\odot}\right)}\right.\\
&&\hspace{-3mm}\left. +\beta_{\rm 2,bulge}\left[\log{\left(\frac{M_{{\rm bulge},i}}{5\times10^{10}M_\odot}\right)}\right]^2\right]. 
\end{eqnarray} 
We performed Monte-Carlo simulations as before and find that the
estimator described in equation~(\ref{LH1_bulge}) leads to a slightly
biased estimate for the true values $\beta_{\rm bulge,true}$ and
$\beta_{\rm 2,bulge,true}$. We find biases of $\Delta\beta_{\rm
bulge}=0.04$ and $\Delta\beta_{\rm 2,bulge}= 0.035$.

In figure~\ref{fig7} we show the a-posteriori marginalised probability
distributions for the parameters $\beta_{\rm bulge}$, $\beta_{\rm
2,bulge}$ and $\delta_{\rm bulge}$ computed using the likelihood
equation~(\ref{LH1_bulge}) and corrected for bias. In the central row
we show joint distributions for $\beta_{\rm 2,bulge}$ and $\beta_{\rm
bulge}$ (left) and for $\delta_{\rm bulge}$ and $\beta_{\rm bulge}$
(right). The contours (dark lines) refer to 0.036, 0.14, 0.26 and 0.64
of the peak height corresponding to the 4, 3, 2, and 1-sigma limits of
a Gaussian distribution. In the bottom rows we show
differential (solid lines) and cumulative (dashed lines) distributions
for $\beta_{\rm bulge}$ (left) and $\beta_{\rm 2,bulge}$ (right). The
vertical dotted lines show the variance. We find that the marginalised
distributions imply $\beta_{\rm bulge}=1.15\pm0.19$, $\beta_{\rm
2,bulge}=0.12\pm0.14$ and an intrinsic scatter of
$\delta=0.41\pm0.07$. The normalisation is $\alpha_{\rm bulge}=8.05\pm0.1$. The
best fit relation has a positive log-quadratic term. However this term is only
required by the data at the 1-sigma level. Haering
\& Rix~(2004) suggested an upper limit on the intrinsic scatter of
$\sim0.3$ dex, though the definition of this scatter was not
specified. We have included intrinsic scatter self consistently in the
analysis and find that the intrinsic scatter in SMBH mass at fixed
bulge mass is $\delta_{\rm bulge}\sim0.4$ dex. This scatter is
larger than the quoted uncertainties on either SMBH or
bulge mass. The scatter in the $M_{\rm bh}-M_{\rm bulge}$ relation is
also $\sim$50\% larger than the scatter in the $M_{\rm bh}-\sigma$
relation.

The best fit log-quadratic relation is plotted over the data (solid
line) in the upper left panel of figure~\ref{fig7} together with
curves showing the level of intrinsic scatter (dotted lines). In the
upper right panel of figure~\ref{fig7} we plot the resulting
residuals, together with horizontal dotted lines showing the value of
the best fit intrinsic scatter. Inspection of the residuals does not
indicate any systematic trend of residuals as a function of $M_{\rm bulge}$.

\section{Discussion}
\label{discussion}

To quantify the extent of the departure of the Local sample from a
log-linear relation we have plotted the logarithmic derivatives
$d\log{M_{\rm bh}}/d\log{\sigma}$ and $d\log{M_{\rm bh}}/d\log{M_{\rm
bulge}}$ of the $M_{\rm bh}-\sigma$ and $M_{\rm bh}-M_{\rm bulge}$
relations as a function of $\sigma$ and $M_{\rm bulge}$ in
figure~\ref{fig8}. The solid lines show the slopes of the best fit
relations. The dashed lines show the slope of relations at the edges of
the 1-sigma error ellipsoids in $\beta-\beta_2$ (left panel) or
$\beta_{\rm bulge}-\beta_{\rm 2,bulge}$ (right panel) space, while the
grey stripes show the 1-sigma range for the slope of the log-linear
fits from T02 and Haering \& Rix~(2004) respectively. The figure shows
that in both cases the logarithmic slope of the log-quadratic relation
varies substantially more over the range of $\sigma$ or $M_{\rm
bulge}$ in the SMBH sample than the statistical uncertainty in the
slope $\beta$ of a log-linear fit. This fact has contributed to
statistical bias and inconsistent results when comparing the
log-linear slopes of the $M_{\rm bh}-\sigma$ relation between
different samples of SMBHs. In the remainder of this section we discuss 
various implications of a log-quadratic $M_{\rm bh}-\sigma$ relation.

\subsection{The most massive SMBH}
The possibility of a log-quadratic $M_{\rm bh}$-$\sigma$ relation
leads to several interesting consequences. The first concerns the
velocity dispersion of galaxies in which the most massive SMBHs
reside. The mass of the largest SMBHs in the Sloan Digital Sky Survey
(SDSS) survey volume was discussed in Wyithe \& Loeb~(2003). We repeat
part of the discussion here. Consider a SMBH of mass $M_{\rm
bh}=3\times10^9M_\odot$. We can estimate the co-moving density of
black-holes of this mass from the observed quasar luminosity
function and an estimate of quasar lifetime. If we assume Eddington
accretion, then at the peak of quasar activity, quasars powered by
$M_{\rm bh}\ga3\times10^9M_\odot$ SMBHs had a comoving density of
$\Psi\sim50$Gpc$^{-3}$. Wyithe \& Loeb~(2003) found that the number of
quasars relative to the number of dark-matter halos at $z\sim3$
implies a duty cycle for quasars of $\sim0.005$. The comoving density
of SMBHs with masses in excess of $M_{\rm bh}\sim3\times10^9M_\odot$
at $z\sim3$ was therefore $\sim10^4$Gpc$^{-3}$. As this corresponds to
the peak of SMBH growth, the SMBH density should match the value
observed today. This density implies that the nearest SMBH of
$\sim3\times10^9M_\odot$ should be at a distance $d_{\rm bh}
\sim 30$Mpc which is comparable to the distance
of M87, a galaxy known to possess a SMBH of this mass (Ford et
al. 1994). What is the most massive SMBH that can be detected
dynamically in the SDSS? The SDSS probes a volume of $\sim1$Gpc$^3$
out to a distance $\sim30$ times that of M87. At the peak of quasar
activity, the density of the brightest quasars implies that there
should be $\sim100$ SMBHs with masses greater than
$\sim3\times10^{10}M_\odot$ per Gpc$^3$, the nearest of which will be
at a distance $d_{\rm bh} \sim 130$Mpc, or 5 times the distance to
M87. The radius of gravitational influence of the SMBH scales as
$M_{\rm bh}/\sigma^2$. We therefore find that for the nearest
$3\times10^9M_\odot$ and $3\times10^{10}M_\odot$ SMBHs, the angular
radius of influence should be similar. Thus the dynamical signature of
the nearest $\sim3\times10^{10}M_\odot$ SMBHs on their stellar host
should be detectable.

\begin{figure}
\vspace*{60mm}
\includegraphics{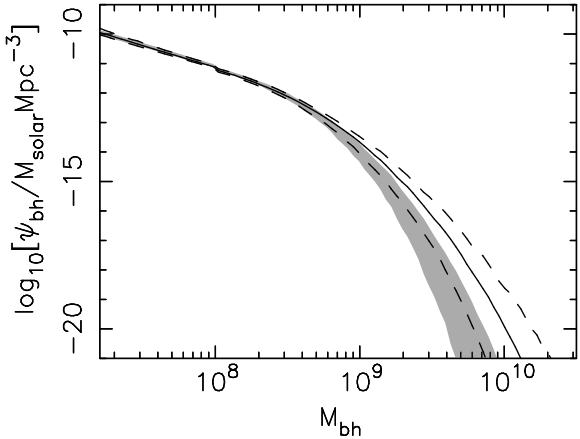}
\caption{\label{fig9} Mass-functions for SMBHs. Two mass
functions are shown. The grey stripe and dashed lines represent the
ranges of uncertainty in the estimate of the mass-function assuming
the log-linear  and log-quadratic relations respectively. The
ranges at fixed $M_{\rm bh}$ in each case correspond to the 16th and
84th percentiles. The solid line shows the most likely estimate for
the mass-function corresponding to the log-quadratic $M_{\rm
bh}-\sigma$ relation.}
\end{figure}

In order to find these most massive SMBHs one needs to identify the
galaxies in which they reside. We estimate the value of $\sigma$
implied for SMBHs of mass $\sim10^{10}M_\odot$ using the mean $M_{\rm
bh}-\sigma$ relation. If one adopts the mean log-linear relation
(T02), then SMBHs with masses in excess of $10^{10}M_\odot$ should
reside in galaxies with $\sigma\ga600$km/s (see
figure~\ref{fig11}). However no such galaxies exist is the SDSS
(e.g. Sheth et al.~2003), where largest values of galaxy velocity
dispersion are found to be $\sigma\sim400$km$\,$s$^{-1}$. The mean
log-quadratic relation reaches $M_{\rm bh}\sim10^{10}M_\odot$ at a
more modest but still unobserved velocity dispersion,
$\sigma\sim500$km/s. However the intrinsic scatter in the $M_{\rm
bh}-\sigma$ relation is $\delta\sim0.3$ dex. We find that SMBHs of
mass $\sim10^{10}M_\odot$ differ by $\la2\delta$ from the mean
log-quadratic relation at $\sigma\sim400$km/s.  The most massive SMBHs
with masses of $\sim10^{10}M_\odot$ inferred from quasars at $z\sim3$
should therefore exist in the SDSS and would lie at $\sim2\delta$
above the extrapolated mean log-quadratic relation for galaxies with
the largest measured velocity dispersions of $\sigma=400$km/s. This
helps to reconcile the number of luminous quasars observed at $z\sim3$
with both the local $M_{\rm bh}-\sigma$ relation and the lack of super
massive galaxies.

\begin{figure*}
\vspace*{60mm}
\includegraphics{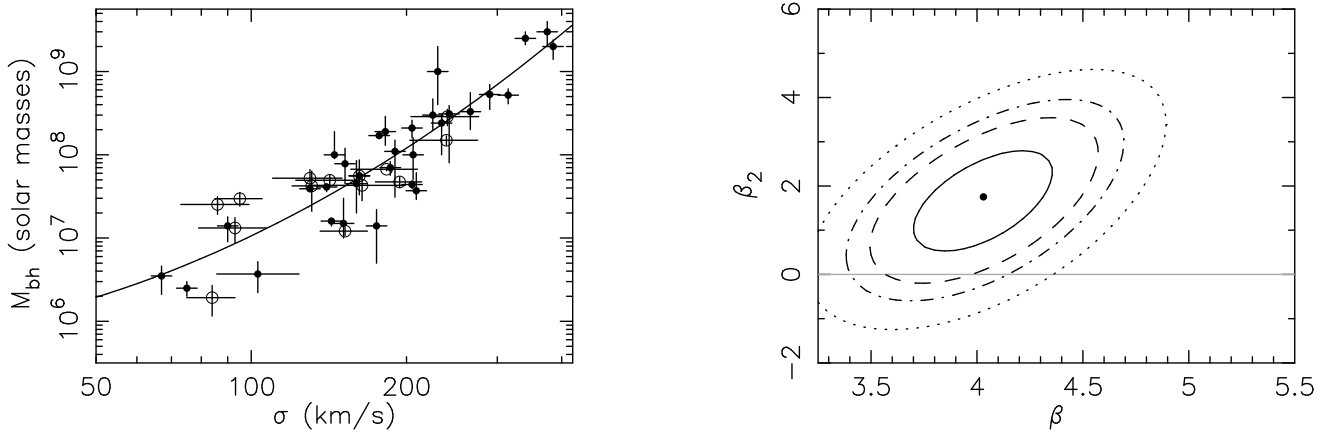}
\caption{\label{fig10} {\em Left}: The Local sample of SMBHs (solid dots) 
and the sample of SMBHs determined from reverberation mapping studies
(with $\langle f_{\rm v}\rangle=5.5$; Onken et al.~2004; open circles). {\em Right:}
The resulting contours of $\chi^2$ (minimised over $\alpha$) of
$\beta$ and $\beta_2$. Projections of the contours correspond to the 1, 1.5, 2 and
2.5-sigma limits on individual parameters. The point shows the most likely solution.  }
\end{figure*}

\subsection{The mass-function of SMBHs}
A positive log-quadratic term has a significant effect on the upper
end of the SMBH mass-function. In figure~\ref{fig9} we show the SMBH
mass-function computed by combining the $M_{\rm bh}-\sigma$ relation
with the velocity dispersion function of Sheth et al.~(2003). Two mass
functions are shown. The grey stripe and dashed lines represent the
ranges of uncertainty in the estimate of the mass-function obtained assuming
log-linear and log-quadratic $M_{\rm bh}-\sigma$ relations
respectively.  The ranges at fixed $M_{\rm bh}$ in each case
correspond to the 16th and 84th percentiles, and were computed taking
into account Gaussian uncertainties in both the parameters of the $M_{\rm
bh}-\sigma$ relation and the velocity-dispersion function. The solid
line shows the most likely estimate for the mass-function
corresponding to the log-quadratic $M_{\rm bh}-\sigma$
relation. Figure~\ref{fig9} shows that inclusion of a log-quadratic
term in the $M_{\rm bh}-\sigma$ relation results in an estimate for
the number density of SMBHs with masses of $\sim10^{10}M_\odot$ that
is several orders of magnitude larger than inferred from a log-linear
relation.  We have estimated the mass density (for a Hubbles constant
of $71$km$\,$s$^{-1}\,$Mpc$^{-1}$) of SMBHs in the local universe. For
a log-linear $M_{\rm bh}-\sigma$ relation we find a density of
$\rho=(2.30\pm0.45)\times10^{5}M_\odot\,$Mpc$^{-3}$, while for a
log-quadratic $M_{\rm bh}-\sigma$ relation we obtain
$\rho=(2.35\pm0.45)\times10^{5}M_\odot\,$Mpc$^{-3}$.  Hence the
addition of a log quadratic term does not effect estimates of the
total mass density which is dominated by SMBHs in the range
$10^7-10^9M_\odot$ (Aller \& Richstone~2002).

\subsection{Inclusion of SMBHs from reverberation mapping studies}

The sample of kinematically detected SMBHs will not grow by a large
factor in the foreseeable future (Ferrarese~2003). Progress in
understanding the statistical properties of the SMBH population may
instead come via estimates of SMBH masses that are based on
reverberation mapping studies (Gebhardt et al.~2000b; Ferrarese et
al.~2001; Onken et al.~2004; Peterson et al.~2004; Nelson et
al~2004). Onken et al.~(2004) have presented a sample of 14 SMBHs with
virial masses ($M_{\rm bh,vir}$) determined via reverberation mapping
studies, in galaxies with measured velocity dispersions. Reverberation
masses are determined up to a geometric factor $f$, hence the SMBH
mass is $M_{\rm bh}=f\times M_{\rm bh,vir}$. Onken et al.~(2004) have
determined the average value $\langle f\rangle$ by comparing the
reverberation masses of these 14 SMBHs, to the masses of SMBHs in the
Local sample using a log-linear $M_{\rm bh}-\sigma$ relation. They
found $\langle f\rangle\sim5.5$ over a range of values for the slope
$\beta$ that include $\beta=4.58$ (Ferrarese~2002) and $\beta=4.02$
(T02). Assuming this scaling factor we can combine the 14 galaxies
used to obtain $\langle f\rangle$ in Onken et al.~(2004) with the
Local sample of SMBHs to form a sample of 45 galaxies.

The left panel of figure~\ref{fig10} shows the $M_{\rm bh}-\sigma$
relation for the Local sample (solid dots), and its comparison to the
AGN SMBH masses and velocity dispersions (open circles) presented in
Onken et al.~(2004). We have repeated the $\chi^2$ minimisation using
equation~(\ref{fit_quad}) for the combined sample of $N_{\rm g}=45$
SMBHs. In the right-hand panel we show the 1, 1.5, 2 and 2.5-sigma
ellipsoids for $\beta$ and $\beta_2$. The most likely solution has
$\beta=4.05$ and $\beta_2=1.8$ with 1-sigma uncertainties of 0.3 and
1.0 respectively. The resulting best-fit $M_{\rm bh}-\sigma$ relation
is over-plotted in the left panel (solid line). This solution is very
close to the one obtained using only SMBHs in the Local sample
($\beta=4.3\pm0.35$, $\beta_2=1.9\pm1.4$). The intrinsic scatter
required for a reduced $\chi^2$ of unity in this fit was
$\epsilon_{\rm intrins}=0.25$ dex, indicating that the addition of the
AGN SMBHs does not increase the scatter of the $M_{\rm bh}-\sigma$
relation. The addition of AGN SMBHs results in a fit with a
log-quadratic term which is non-zero at around the 1.5-sigma level. The sample
of AGN SMBHs with masses determined from reverberation mapping
therefore strongly supports a log-quadratic $M_{\rm bh}-\sigma$
relation. This is despite the fact that the normalization of the
reverberation SMBH masses was determined via the assumption of a
log-linear relation.

\begin{figure*}
\vspace*{130mm}
\includegraphics{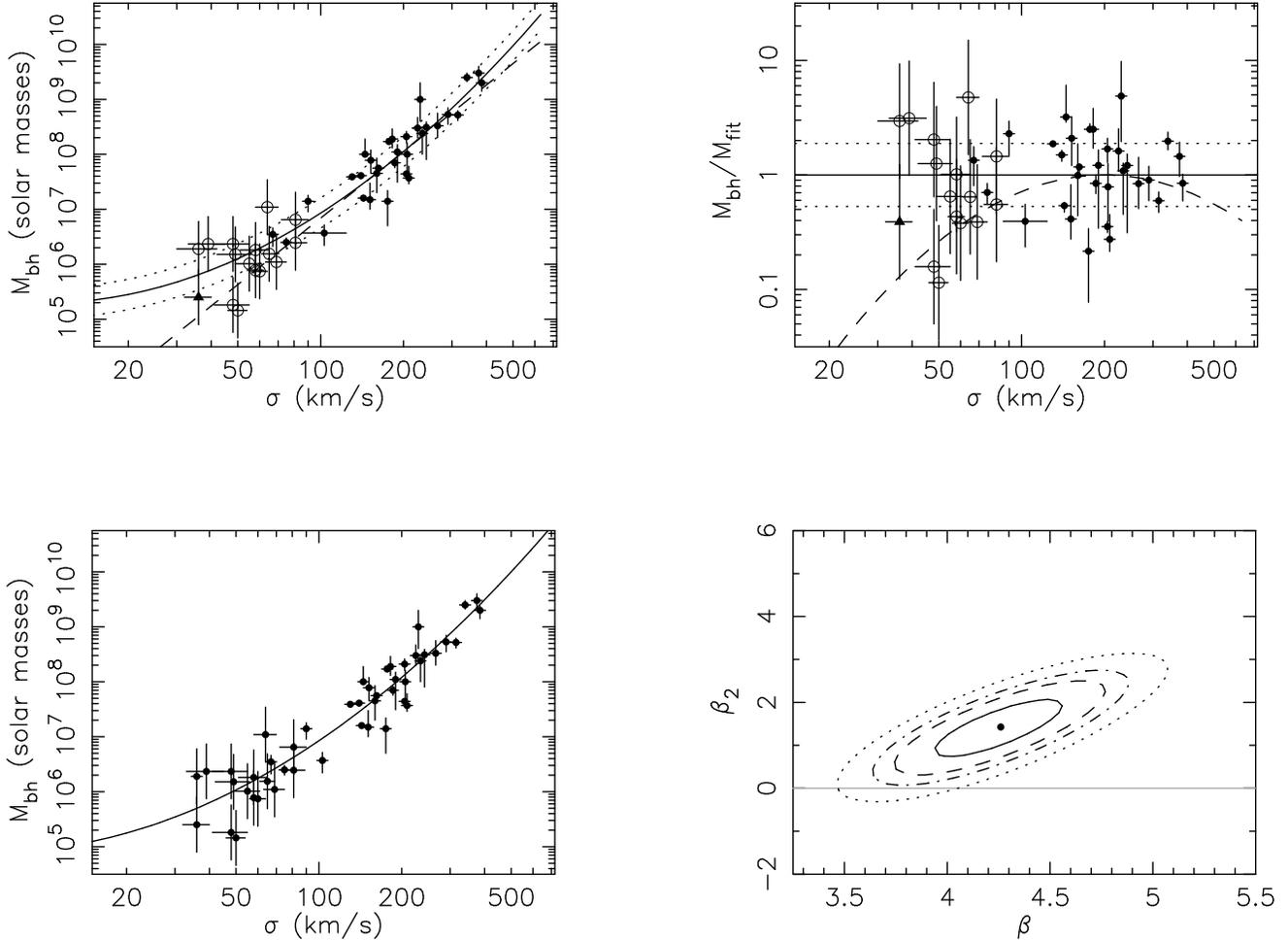}
\caption{\label{fig11} {\em Upper Left}: The best fit log-quadratic relation (determined from the Bayesian analysis of the Local sample) extrapolated to high and low $\sigma$ (solid line), and the level of intrinsic scatter around the best-fit relation (dotted lines). Also shown (dashed line) is the log-linear relation from T02. The points from the Local sample are shown (solid dots). In addition, the open circles represent SMBHs with masses estimated using AGN identified in SDSS (Barth, Greene \& Ho.~2004). The triangle represents POX52 (Barth, Ho, Rutledge \& Sargent~2004). {\em Upper Right}: The corresponding residuals relative to the log-quadratic relation, with dotted lines showing the best fit intrinsic scatter. The dashed line shows the additional residual of the log-quadratic relation relative to the log-linear relation. {\em Lower Left:} The combined sample of Local and AGN SMBHs with the resulting best-fit log-quadratic relation determined from a $\chi^2$ analysis. {\em Lower Right:} The resulting contours of $\chi^2$ (minimised over $\alpha$) of $\beta$ and $\beta_2$. Projections of the contours correspond to the 1, 1.5, 2 and 2.5-sigma limits on individual parameters. The point shows the most likely solution.     }
\end{figure*}

\subsection{The low mass $M_{\rm bh}-\sigma$ relation}
There are no SMBHs in the Local sample with masses below
$\sim2\times10^6M_\odot$. Attempts to extend the $M_{\rm bh}-\sigma$
relation to lower masses through dynamical observations have centered
on two cases, M33 (Merritt, Ferrarese \& Joseph~2001; Gebhardt et
al. 2001), and NGC205 (Valluri, Ferrarese, Merritt \&
Joseph~2005). Only upper limits have been obtained in these
systems. However we can compare the log-quadratic fit in the
low-$\sigma$ regime with SMBH masses obtained from single epoch virial
estimates based on observations of AGN. The upper panel of
figure~\ref{fig11} shows the extrapolation of the best fit
log-quadratic relation for the Local sample, and its comparison to the
AGN SMBH masses and velocity dispersions presented in Barth, Greene \&
Ho~(2004), as well as the point for POX52 (Barth, Ho, Rutledge \&
Sargent~2004). The AGN SMBH masses in this sample were estimated using
the single epoch virial mass calibration of Onken et al.~(2004).  The
log-quadratic fit derived using the Local sample appears to describe
the SMBH masses in these AGN much better than the log-linear fit (see
right-hand panel of Figure~\ref{fig11}). The residuals are
symmetrically placed around the best fit log-quadratic relation. On
the other hand the dashed line in the right hand panel of
figure~\ref{fig11} shows the additional residual of the log-quadratic
relation relative to the log-linear relation. The residuals of the AGN
SMBHs relative to the log-linear relation are systematically positive,
indicating that the log-linear relation provides a poor description.

The single epoch virial SMBH masses for the AGN have been estimated in
a regime where the technique is not directly calibrated. One must
therefore be cautious comparing the AGN and local samples since the
AGN SMBH mass estimates may contain some unknown systematic
uncertainty relative to the dynamical estimates of the Local sample
(Barth, Greene \& Ho~2005 did not perform a fit to the combined sample
for this reason). Moreover it is possible that the AGN SMBH sample is
biased towards large SMBH masses (Barth, Greene \& Ho~2005).
Never-the-less it is instructive to combine the Local and AGN SMBH
samples and investigate the resulting log-quadratic $M_{\rm
bh}-\sigma$ relation. We have repeated the $\chi^2$ minimisation using
equation~(\ref{fit_quad}) for the combined sample of $N_{\rm g}=47$
SMBHs. In the lower-left panel of figure~\ref{fig11} we show the
combined sample with the resulting best-fit $M_{\rm bh}-\sigma$
relation over-plotted (solid line). The intrinsic scatter required for
a reduced $\chi^2$ of unity in this fit was $\epsilon_{\rm
intrins}=0.22$ dex. In the lower-right panel of figure~\ref{fig11} we
show the 1, 1.5, 2 and 2.5-sigma ellipsoids for $\beta$ and $\beta_2$. The
most likely solution has $\beta=4.25$ and $\beta_2=1.3$ with 1-sigma
uncertainties of 0.3 and 0.7 respectively. This solution is very close
to the one obtained using only SMBHs in the Local sample
($\beta=4.3\pm0.35$, $\beta_2=1.9\pm1.1$ and $\epsilon_{\rm
intrins}=0.25$ dex), indicating that the AGN SMBHs follow the same
log-quadratic $M_{\rm bh}-\sigma$ relation as the Local
sample. However the addition of low-mass AGN SMBHs results in a fit
with a log-quadratic term which is non-zero at greater than the 2-sigma level. The
sample of AGN SMBHs therefore support a log-quadratic $M_{\rm
bh}-\sigma$ relation. The log-quadratic relation describes the masses
of SMBHs in galaxies with velocity dispersions ranging from
25-400km$\,$s$^{-1}$. If one extrapolates this to lower $\sigma$ then
the log-quadratic relation offers the tantalising prediction that
there is a minimum SMBH mass of $\sim10^5M_\odot$, and that this
minimum SMBH mass resides in bulges with
$\sigma\sim10$km$\,$s$^{-1}$. Interestingly, this velocity dispersion
corresponds to the minimum value of the virial velocity for a
dark-matter halo within which gas accreted from the IGM can cool via
atomic transitions in hydrogen.

\subsection{The relation between $\beta$ and $\beta_{\rm bulge}$}
Finally we would like to make the following point regarding the
relationship between the power-law slope of the $M_{\rm bh}-\sigma$
and $M_{\rm bh}-M_{\rm bulge}$ relations. An estimate of the bulge
mass may be made using the virial mass, $M_{\rm bulge}\sim M_{\rm
bulge,vir}\propto\sigma^2R_{\rm e}$, where $R_{\rm e}$ is the scale
radius of the bulge. The projection of the fundamental plane yields a
relation between radius and velocity dispersion of $R_{\rm
e}\propto\sigma^{1.5}$ (Bernardi et al.~2003).  Consider the power-law
slope at the mean velocity of the galaxy sample
($\sigma\sim200$km/s). We have $M_{\rm bh}\propto (M_{\rm
bulge})^{\beta_{\rm bulge}}$, which implies $M_{\rm bh}\propto
\sigma^{3.5\beta_{\rm bulge}}$, or $\beta=3.5\beta_{\rm bulge}$. This
is in excellent agreement with the best fit parameters of
$\beta\sim4.2$ and $\beta_{\rm bulge}=1.15$ derived in this
paper. Furthermore, a value of $\beta=3.5$ for the $M_{\rm bh}-\sigma$
relation is ruled out at $\sim80\%$ significance. If it is to be
consistent with the $M_{\rm bh}-\sigma$ relation, the $M_{\rm
bh}-M_{\rm bulge}$ relation should therefore be steeper than linear
for galaxies with $\sigma\sim200$km/s. Given the relation between
$\sigma$ and the virial mass, we would also expect the $M_{\rm
bh}-M_{\rm bulge}$ relation to have a positive quadratic component,
given a positive quadratic component in the $M_{\rm bh}-\sigma$
relation.

\section{Conclusion}
\label{conclusion}

Using the Local sample of SMBHs we have demonstrated that the
parameter describing a log-linear fit to the $M_{\rm bh}-\sigma$
relation is sensitive to the inclusion of individual galaxies at a
level larger than the statistical uncertainty. This indicates that a
log-linear $M_{\rm bh}-\sigma$ relation does not provide a good
description of the Local sample of SMBHs. We expand a general relation
between $\log(M_{\rm bh})$ and
$\log(\sigma/200\mbox{km}\,\mbox{s}^{-1})$ to second order, and fit
the data using the resulting log-quadratic relation instead. We find
that a log-quadratic relation provides a substantially better fit to
the Local sample of SMBH masses and velocity
dispersions. Moreover unlike the log-linear relation, the parameters
describing the log-quadratic relation are not systematically dependent
on the inclusion of individual galaxies in the sample.

After allowing for a second-order term in the $M_{\rm bh}-\sigma$
relation we find an unbiased estimate for the slope of the Local
sample at $\sigma=200$km/s to be $\beta=4.2\pm0.37$. This value is
slightly (2/3-sigma) larger than previous estimates for this
sample. However the logarithmic slope $d\log{M_{\rm
bh}}/d\log{\sigma}$ of the best fit log-quadratic relation varies
substantially, from 2.7-5.1 over the velocity range of the Local
sample.  The coefficient of the second order term $\beta_2=1.6\pm1.3$
is greater than zero at the 90\% level. This indicates that with 80\%
confidence the Local sample describes an $M_{\rm bh}-\sigma$ relation
that does not follow a single powerlaw between $\sim70$km/s and
$\sim380$km/s. We have tested the sensitivity of this conclusion to
different sub-samples of SMBH masses determined via different techniques,
as well as SMBHs with and without resolved spheres of influence. The
log-quadratic fit implies a non-linear contribution to the $M_{\rm
bh}-\sigma$ relation in a sub-sample of galaxies that contain only
SMBHs whose spheres of influence were resolved, in a sub-sample
containing only SMBHs with masses determined through stellar dynamics,
and in a sub-sample containing only SMBHs with masses determined only
by non stellar dynamical methods. Moreover the non-linearity is
present in each sub-sample whether the central or effective velocity
dispersion is used as the independent variable.

SMBH masses in active galaxies estimated via reverberation mapping
offer an avenue to increase the SMBH sample for study of the $M_{\rm
bh}-\sigma$ relation. We find that the combination of the 14 galaxies
with reverberation SMBH masses and measured velocity dispersions
(Onken et al.~2004) with the Local SMBH sample leads to a
log-quadratic relation with the same best fit as the Local sample
alone.

The $M_{\rm bh}-\sigma$ relation can be extended to lower masses through the
inclusion of single epoch virial estimates of SMBH masses based on
observations of AGN (Barth, Greene \& Ho~2004). In a $\chi^2$ analysis
the best-fit log-quadratic $M_{\rm bh}-\sigma$ relation is unchanged
by the addition of a sample of 16 low mass SMBHs. However the
uncertainty in $\beta_2$ is reduced. We find that in the
absence of a systematic error in the normalisation of SMBH masses
between the Local and AGN samples, the $M_{\rm bh}-\sigma$ relation
described by the combined sample deviates from a power-law at greater than the
2-sigma level. The best-fit log-quadratic relation predicts a minimum
mass for SMBHs in galaxies of $\sim10^5M_\odot$, which should reside
in bulges with $\sigma\sim10$km$\,$s$^{-1}$.

A log-quadratic $M_{\rm bh}-\sigma$ relation has important
implications for SMBH demography. In particular, estimates of the
local SMBH mass-function that utilise the log-quadratic $M_{\rm
bh}-\sigma$ relation describe densities of SMBHs with $M_{\rm
bh}\ga10^9M_\odot$ that are orders of magnitude larger than expected
for a log-linear $M_{\rm bh}-\sigma$ relation.  In addition the
departure from a power-law should provide important clues regarding
the astrophysics responsible for the $M_{\rm bh}-\sigma$ relation. For
example one recent model describing the effects of radiative feedback
on SMBH growth predicts departure from a power-law relation, including
steepening of the relation at large $\sigma$ (Saznov et al.~2005).

We have also applied our analysis to the relation between SMBH and
bulge mass using the sample described in Haering \& Rix~(2004). We
find evidence for a log-quadratic term in the $M_{\rm bh}-M_{\rm
bulge}$ relation (at the 1-sigma level), with $\beta_{\rm
bulge}=1.15\pm0.19$ and $\beta_{\rm 2,bulge}=0.12\pm0.14$. We find an
intrinsic scatter of $\delta_{\rm bulge}=0.41\pm0.07$ dex. Since we
find the intrinsic scatter in the $M_{\rm bh}-\sigma$ relation to
be $\delta=0.28\pm0.04$ dex, there is $\sim50\%$ more scatter in the
SMBH mass at fixed bulge mass than at fixed velocity dispersion.

The sample of kinematically detected SMBHs will not grow by a large
factor in the foreseeable future (Ferrarese~2003). Progress in
understanding the statistical properties of the SMBH population will
instead come via estimates of SMBH masses that are based on
reverberation mapping studies. The increased sample of SMBHs will
offer the possibility of more clearly defining the local $M_{\rm
bh}-\sigma$ relation (Gebhardt et al.~2000b; Ferrarese et al.~2001;
Onken et al.~2004; Peterson et al.~2004; Nelson et al~2004), and of
extending its study to high redshift (Shields et al.~2003).

\section*{Acknowledgments}

The author would like to thank Scott Tremaine for some very helpful comments
including the suggestion of introducing a log-quadratic term, as well as Christian
Knigge and Alister Graham for pointing out errors in the original version of the
paper. The
author acknowledges the support of the Australian Research Council.

\begin{appendix}

\section{Comparison of parameters for different SMBH samples}
\label{app1}

\begin{figure*}
\vspace*{130mm}
\includegraphics{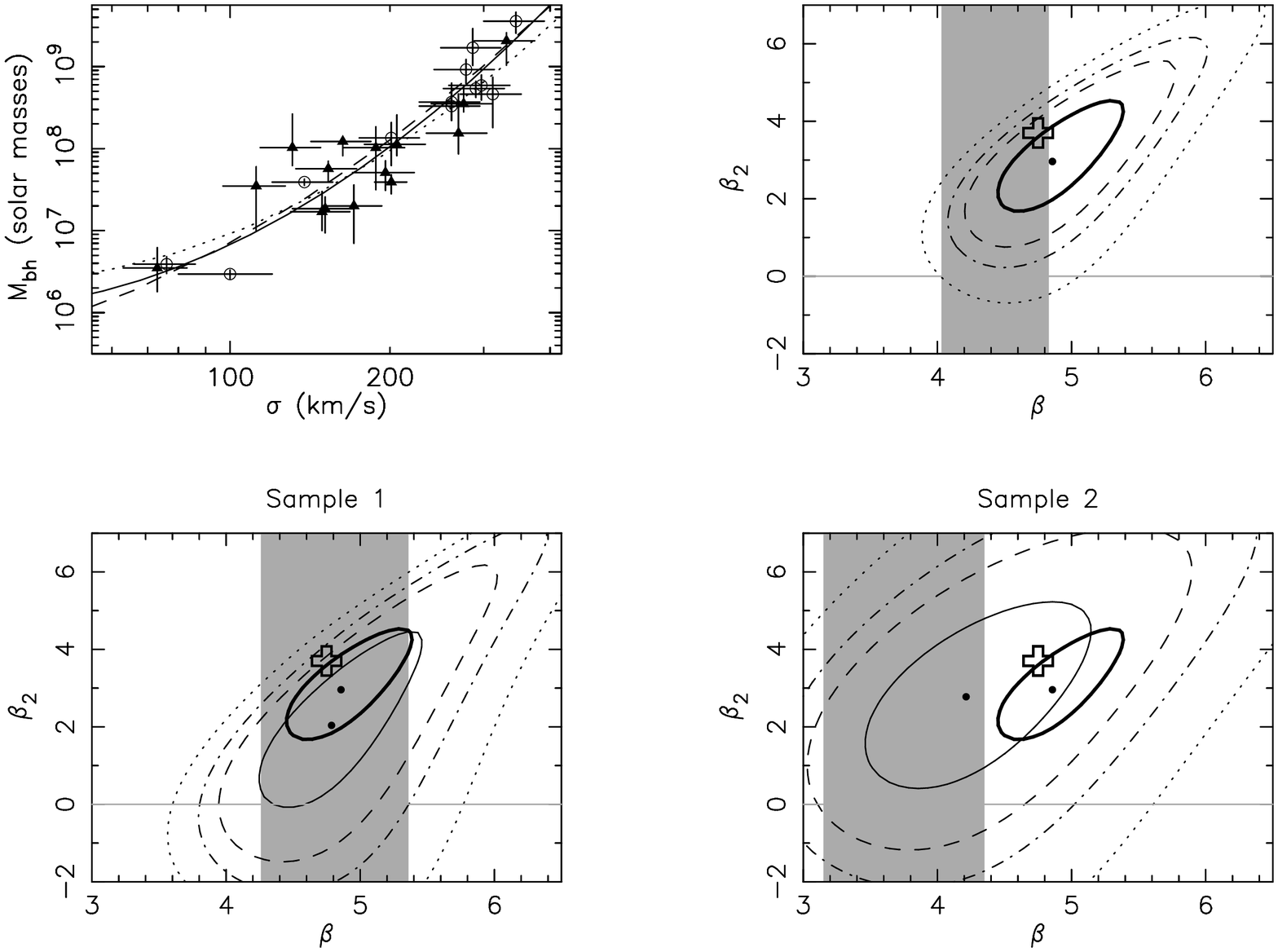}
\caption{\label{fig12} Log-quadratic fits to the samples of Ferrarese \& Merritt~(2000, sample~1), the additional galaxies added by Gebhardt et al.~(2000, sample~2), as well as a combined sample (upper right panel) as described in Merritt \& Ferrarese~(2001). Projections of the contours represent the 1, 1.5, 2 and 2.5-sigma uncertainties on individual parameters. The 1-sigma error ellipse for the combined sample is repeated in the lower panels. In this figure the correlation is between SMBH mass and the central velocity dispersion (Ferrarese \& Merritt~2000). The grey region shows the corresponding linear fit value for $\beta$ from Merritt \& Ferrarese~(2001). The upper left panel shows the three best fits along with the data. Sample-1 is represented by open circles, and sample-2 by triangles. The large cross shows the point $\beta=4.9$, $\beta_2=3.75$ which lies inside the 1-sigma error ellipses of all three samples in both the cases where central and effective velocity dispersion are used as the independent variable.}
\end{figure*}

\begin{figure*}
\vspace*{130mm}
\includegraphics{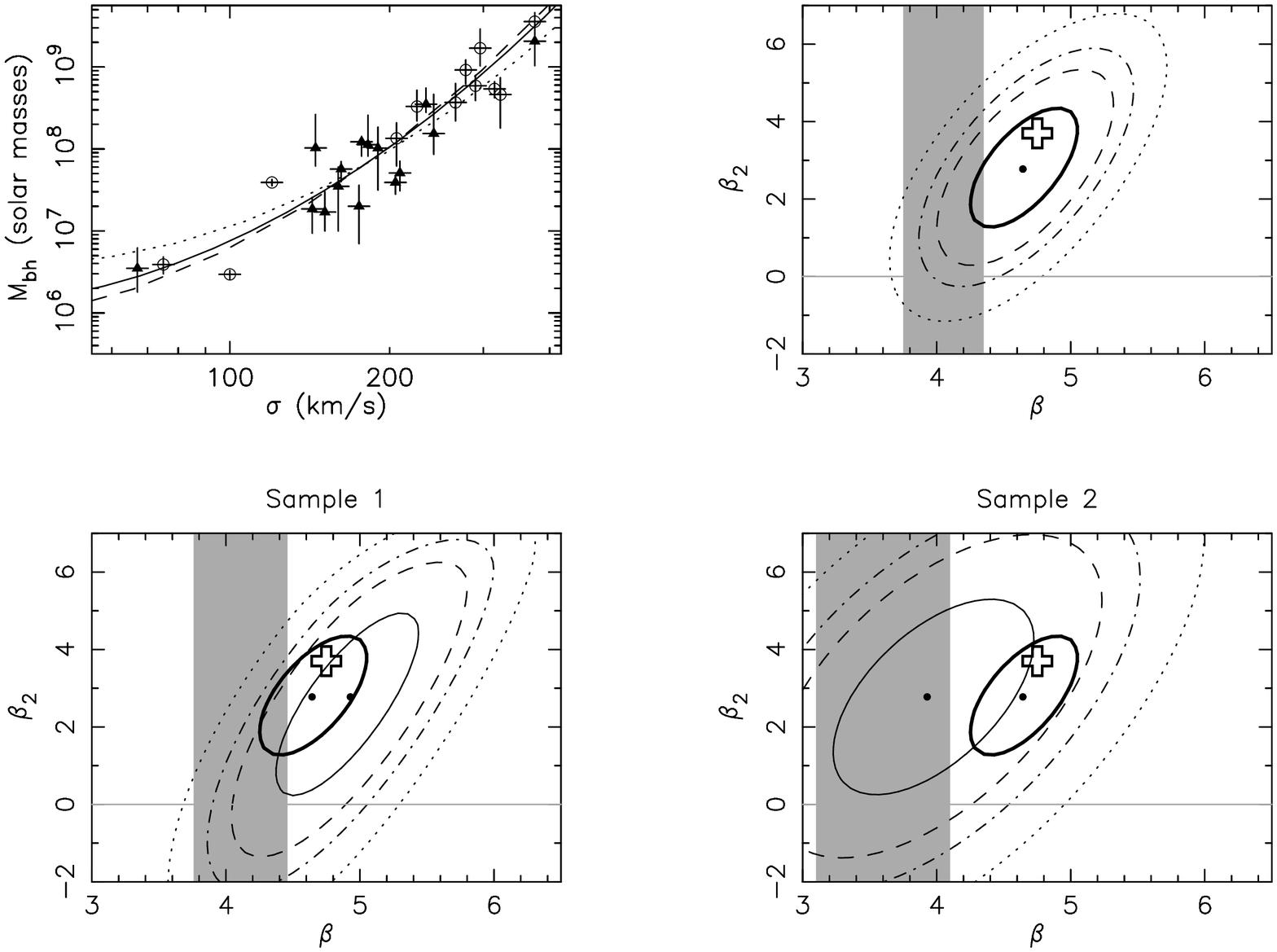}
\caption{\label{fig13}  Log-quadratic fits to the samples of Ferrarese \& Merritt~(2000, sample~1), the additional galaxies added by Gebhardt et al.~(2000, sample~2), as well as a combined sample (upper right panel) as described in Merritt \& Ferrarese~(2001). Projections of the contours represent the 1, 1.5, 2 and 2.5-sigma uncertainties on individual parameters. The 1-sigma error ellipse for the combined sample is repeated in the lower panels. In this figure the correlation is between SMBH mass and the effective velocity dispersion (Gebhardt et al.~2000). The grey region shows the corresponding linear fit value for $\beta$ from Merritt \& Ferrarese~(2001). The upper left panel shows the three best fits along with the data. Sample-1 is represented by open circles, and sample-2 by triangles. The large cross shows the point $\beta=4.9$, $\beta_2=3.75$ which lies inside the 1-sigma error ellipses of all three samples in both the cases where central and effective velocity dispersion are used as the independent variable.}
\end{figure*}

Following the discovery of the $M_{\rm bh}-\sigma$ relation (Gebhardt
et al.~2000, Ferrarese \& Merritt~2000), there has been debate in the
literature among the different groups regarding inconsistencies in the
value of the measured slope. These differences have been attributed to
several causes including use of biased statistics (Merritt \&
Ferrarese~2001) and systematic differences in the velocity dispersions
used for the host galaxies (T02).  We have shown that the slope
$\beta$ of a log-linear relation is substantially changed by the
presence or absence of one or a couple of the smallest galaxies in the
sample. Given the different samples of velocity dispersions that have
been employed by the different groups, the bias introduced by the
assumption of a log-linear relation may therefore be more or less
significant for different samples.

Merritt \& Ferrarese~(2001) presented an analysis of the log-linear
relation for three different samples. The first two samples were drawn
from Ferrarese \& Merritt~(2000, sample~1), and from the additional
galaxies presented by Gebhardt et al.~(2000, sample 2). The accuracy
of some of the additional SMBH masses included by Gebhardt et
al.~(2000) has been called into question on the basis that the SMBH
sphere of influence was not resolved (e.g. Merritt \& Ferrarese~2001).
In addition, Merritt \& Ferrarese~(2001) discuss the combined sample
from these two sub-samples. For sample~2 Merritt \& Ferrarese~(2001)
computed aperture corrected central velocity dispersions, so that all
galaxies in the combined sample could be considered when estimating
correlations between SMBH mass and both a central (Ferrarese \&
Merritt~2000) and an effective velocity dispersion (Gebhardt et
al.~2000). Merritt \& Ferrarese~(2001) present fits for the slope
$\beta$ of a log-linear relation to these samples using regression
with bi-variate errors and intrinsic scatter (their label BRS).

We first consider the correlation of SMBH mass with central velocity
dispersion (Ferrarese \& Merritt~2000). Merritt \& Ferrarese~(2001)
find $\beta=4.81\pm0.55$ and $\beta=3.75\pm0.59$ for samples~1 and 2
respectively. For the combined sample they get $\beta=4.43\pm0.39$.
Samples~1 and 2 differ at the 2-sigma level, while the combined sample
leads to a slope that lies between the two. To mimic this analysis for
a log-quadratic relation we minimise the following $\chi^2$ statistic
\begin{equation}
\label{chi2_app}
\chi^2 = \sum_{i=1}^{N_{\rm g}}\frac{(y_i-\alpha - \beta x_i- \beta_2 x_i^2)^2}{\epsilon_{\rm intrins}^2+\epsilon_{yi}^2+(\beta^2+2\beta_2x_i)^2\epsilon_{xi}^2},
\end{equation}
where $y_i$ and $x_i$ are the logarithm of SMBH mass in solar masses
and the logarithm of velocity dispersion in units of 200km/s
respectively. The variables $\epsilon_{xi}$ and $\epsilon_{yi}$ are
the uncertainties in dex for these parameters. The variable
$\epsilon_{\rm intrins}$ is the intrinsic scatter, adjusted to yield a
reduced $\chi^2$ of unity for the best-fit solution. In
figure~\ref{fig12} we show the 1, 1.5, 2 and 2.5-sigma error ellipsoids for
$\beta$ and $\beta_2$ in sample-1, sample-2 and the combined
sample. The 1-sigma ellipsoid for the combined sample is shown in the
other panels for comparison, and the best fit relations are plotted
over the data in the upper left panel. The grey regions represent the
1-sigma range for $\beta$ determined by Merritt \&
Ferrarese~(2001). The ellipsoids show that all three samples suggest a
positive quadratic term is present in the $M_{\rm bh}-\sigma$
relation. The values of $\beta$ determined by Merritt \&
Ferrarese~(2001) are conditional probabilities for $\beta$ given
$\beta_2=0$, and so should correspond to regions where the error
ellipsoids cross the $\beta_2=0$ line. Figures~\ref{fig12} and
\ref{fig13} show this to be the case. The inclusion of a quadratic
term in the relation leads to similar estimates for parameters using
each sample, with $\beta\sim5$ and $\beta_2\sim4$ lying inside the
1-sigma ellipsoids.

We also investigate the correlation between SMBH mass and effective
velocity dispersion (Gebhardt et al.~2000). The values for effective
velocity dispersion $\sigma$ in Gebhardt et al.~(2000) did not include
an estimate of the uncertainty.  In the absence of quoted
uncertainties, Merritt \& Ferrarese~(2001) assumed different values
for the uncertainty in effective velocity dispersion and showed that
increasing the uncertainty lead to steeper best fit slopes, and to
consistency between fits for samples 1 and 2. T02 has since advocated
5\% for the fractional uncertainty in effective velocity
dispersion. We assume 5\% errors in $\sigma$ and adopt
$\sigma=100$km/s for the Milky-Way. In this case Merritt \&
Ferrarese~(2000) found $\beta=3.6\pm0.5$ and $\beta=4.11\pm0.35$ for
samples 1 and 2 respectively. For the combined sample, they found
$\beta=4.05\pm0.3$.  In figure~\ref{fig13} we show the results for the
log-quadratic fit in this case. The results are consistent for a
log-quadratic relation, with $\beta\sim4.6$ and $\beta_2\sim3$ lying
inside the 1-sigma error ellipses of all three samples. Consistency
between the best fit log-quadratic relations for samples~1 and 2
therefore does not require uncertainties in $\sigma$ to be larger than
5\%, the value advocated by the Nuker team (Gebhardt et al.~2003). 

Finally we compare the estimates of $\beta$ and $\beta_2$ for
relations between SMBH mass, and the central or effective velocity
dispersions using the combined sample. The use of an effective
velocity dispersion (with 5\% uncertainty) leads to smaller values of
$\beta$ and $\beta_2$ relative to a fit using central velocity
dispersions. Merritt \& Ferrarese~(2001) found that this difference
could be removed by increasing the uncertainty in $\sigma$ to greater
than 10\%. However a solution with $\beta\sim4.9$ and $\beta_2\sim3.5$
lies inside the 1-sigma error ellipsoids in both cases.  Overall we
find that a solution with $\beta\sim4.75$ and $\beta_2\sim3.7$ (shown by the
large cross in figures~\ref{fig12} and \ref{fig13}) lies on or inside
the 1-sigma error ellipsoids of all three samples using either the
central or the effective velocity dispersions as the independent variable.
Therefore if a log-quadratic $M_{\rm bh}-\sigma$ relation is used we
find that there is no significant disparity between the different
samples.

\section{An updated sample of SMBH mass and central velocity dispersion}
\label{app2}

\begin{figure*}
\vspace*{130mm}
\includegraphics{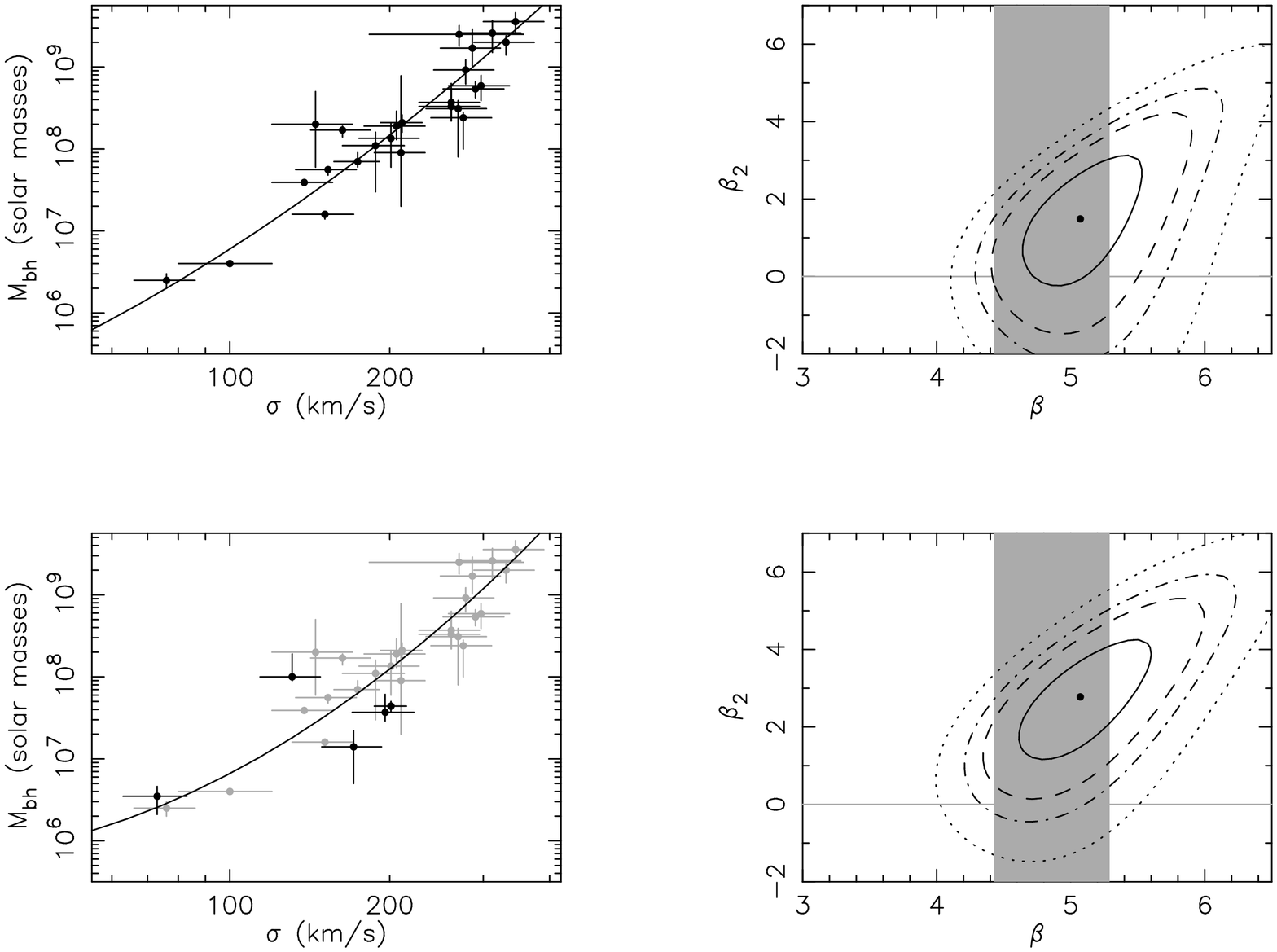}
\caption{\label{fig14} {\em Upper left}: The sample of 25 SMBHs with reliable masses and resolved spheres of influence described in Ferrarese \& Ford~(2004). {\em Upper right:}
The resulting contours of $\chi^2$ (minimised over $\alpha$) of
$\beta$ and $\beta_2$. Projections of the contours correspond to the 1, 1.5, 2 and
2.5-sigma limits on individual parameters. The point shows the most likely solution.  {\em
Lower left and right:} As for the upper row, but including the
additional 5 SMBH masses in table~II of Ferrarese \& Ford~(2004)
classed as reliable but with unresolved spheres of influence (shown as
dark points). The grey regions show the allowed range of $\beta=4.86\pm0.43$ in a log-linear fit to the 25 SMBHs with a resolved sphere of influence (Ferrarese \& Ford~2004).}
\end{figure*}

In a recent review, Ferrarese \& Ford~(2004) have compiled an updated
sample of local SMBHs, listed in their table-II. Of the 38 mass
estimates, they class 8 as unreliable. Of the remaining 30 SMBHs, the
mass observations resolved the sphere of influence in only 25
cases. Ferrarese \& Ford~(2004) used this sample of 25 reliable mass
estimates where the sphere of influence is resolved to generate a
log-linear fit to the $M_{\rm bh}-\sigma$ relation. They find a
log-linear slope of $\beta=4.86\pm0.43$.  We have performed
log-quadratic fits to this updated sample, and show the results in
Figure~\ref{fig14}. In the upper left panel we show the sample of 25 SMBHs
with resolved spheres of influence, and the corresponding best-fit
relation. In the upper right-hand panel we show the 1, 1.5, 2 and 2.5-sigma
error ellipsoids of $\beta$ and $\beta_2$ for this sample. This
sample prefers a log-quadratic relation, with the log-linear relation
lying near the $\Delta\chi^2=1$ contour. The grey regions in
Figure~\ref{fig14} represent the 1-sigma range for $\beta$ determined
by Ferrarese \& Ford~(2004). We see that the contours of $\chi^2$
intercept the $\beta_2=0$ line over a range of slope described by the
log-linear solution.

Of the 5 galaxies with reliable mass-estimates, but unresolved spheres
of influence in Table~II of Ferrarese \& Ford~(2004), 4 have had their
stellar kinematical data modeled using both high-resolution spaced
based data, and lower resolution ground based data (Gebhardt et
al.~2003). Gebhardt et al.~(2003) found that while higher resolution
imaging led to a more precise determination, there was no systematic
trend of the best fit values. We therefore repeat the above analysis
with the inclusion of the 5 galaxies and show the results in the lower
row of Figure~\ref{fig14}. The additional 5 galaxies only change the
slope $\beta$ of a log-linear fit by a small amount (significantly
less than the error bar). Moreover, these 5 SMBH masses are evenly
spread about the best-fit relation, indicating that there is no
systematic bias due to their spheres of influence not being
resolved. However the inclusion of these galaxies significantly
increases the significance of a log-quadratic term in the $M_{\rm
bh}-\sigma$ relation.

\end{appendix}

\label{lastpage}

\end{document}